\documentclass{article}
\usepackage{graphicx} 
\usepackage{mathrsfs}
\usepackage{MnSymbol}
\usepackage{amsthm,amsmath}
\usepackage{amsfonts}
\usepackage{bbm}
\allowdisplaybreaks
\usepackage[letterpaper,top=2cm,bottom=2cm,left=2cm,right=2cm,marginparwidth=1.75cm]{geometry}
\usepackage{authblk}
\usepackage{setspace}
\usepackage{natbib}
\usepackage{booktabs}
\usepackage{lipsum}  
\usepackage{float}
\usepackage{graphicx}
\usepackage{subcaption}
\usepackage{xcolor}
\usepackage{lineno}
\definecolor{c40}{rgb}{0,0.5,1}

\newcommand{\Ind}[2]{\ensuremath{{\mathbbm 1}_{#1}\left(#2\right)}}

\newcommand{\p}[1]{\ensuremath{\left(#1\right)}}
\newcommand{\N}{\ensuremath{\mathbb N}}

\newtheorem{definition}{Definition}
\newtheorem{assumption}{Assumption}
\newtheorem{lemma}{Lemma}

\title{Estimating HIV Cross-sectional Incidence Using Recency Tests from a Non-representative Sample}

\author[1]{Jianan Pan}
\author[1]{Marlena Bannick}
\author[2,3]{Fei Gao \thanks{Correspondence: Fei Gao, Biostatistics, Bioinformatics and Epidemiology Program, Fred Hutchinson Cancer Center, Seattle, WA, USA.
Email: fgao@fredhutch.org}}
\affil[1]{\small Department of Biostatistics, University of Washington, Seattle, Washington, USA}
\affil[2]{Biostatistics, Bioinformatics and Epidemiology Program, Fred Hutchinson Cancer Center, Seattle, Washington, USA
}
\affil[3]{Public Health Sciences Division, Fred Hutchinson Cancer Research Center, Seattle, Washington, USA}
\date{}
\begin{document}
\maketitle
\begin{abstract}
Cross-sectional incidence estimation based on recency testing has become a widely used tool in HIV research. 
Recently, this method has gained prominence in HIV prevention trials to estimate the ``placebo" incidence that participants might experience without preventive treatment. 
The application of this approach faces challenges due to non-representative sampling, as individuals aware of their HIV-positive status may be less likely to participate in screening for an HIV prevention trial. To address this, a recent phase 3 trial excluded individuals based on whether they have had a recent HIV test. To the best of our knowledge, the validity of this approah has yet to be studied.
In our work, we investigate the performance of cross-sectional HIV incidence estimation when excluding individuals based on prior HIV tests in realistic trial settings.
We develop a statistical framework that incorporates an testing-based criterion and possible non-representative sampling. We introduce a metric we call the effective mean duration of recent infection (MDRI) that mathematically quantifies bias in incidence estimation. We conduct an extensive simulation study to evaluate incidence estimator performance under various scenarios.
Our findings reveal that when screening attendance is affected by knowledge of HIV status, incidence estimators become unreliable unless all individuals with recent HIV tests are excluded. Additionally, we identified a trade-off between bias and variability: excluding more individuals reduces bias from non-representative sampling but in many cases increases the variability of incidence estimates.
These findings highlight the need for caution when applying testing-based criteria and emphasize the importance of refining incidence estimation methods to improve the design and evaluation of future HIV prevention trials.

\end{abstract}
\section{Introduction}
Estimating the incidence of HIV based on a single cross-sectional survey has widespread utility. This approach is more convenient and cost-effective than traditional methods for incidence estimation that rely on longitudinal cohort studies \citep{kim2019tracking, busch2010beyond}. \textit{Cross-sectional incidence estimation} is made possible by single or multiple biomarker assays \citep{kassanjee2012new, gao2022statistical} that can differentiate between recent and long-standing HIV infections in HIV-positive individuals \citep{bekker2024twice}.
 used assays include the BED capture enzyme immunoassay (BED CEIA) \citep{mcdougal2006comparison} and the Limiting Antigen Avidity Assay (LAg-Avidity Assay) \citep{lau2022systematic}, which measure HIV-specific IgG levels and HIV-1 antibody avidity, respectively. Based on the properties of these recency tests and under specific assumptions, several incidence estimators have been developed to estimate HIV incidence in target populations with a cross-sectional sample \citep{kaplan1999snapshot, kassanjee2012new, bannick2024enhanced}.

The past decade has seen remarkable progress in developing highly efficacious HIV pre-exposure prophylaxis (PrEP) agents \citep{grant2010preexposure,landovitz2021cabotegravir,bekker2024twice}.
However, the high efficacy of these agents poses challenges for developing new HIV prevention products: it may be unethical to conduct placebo-controlled trials in target populations, and active-controlled trials often require impractically large sample sizes.
Recently, applying cross-sectional incidence estimators to data collected from the screening populations of HIV prevention trials has gained attention as a way to address this challenge. Under certain assumptions, this method can be seen as estimating the HIV incidence  participants would experience without preventive treatment (in other words, a ``placebo'' incidence) \citep{gao2021sample}. The absolute efficacy of the investigational prevention treatment can then be evaluated using this benchmark of background incidence.

A common challenge of using the recency testing method to estimate background incidence is that the individuals being screened for a prevention trial may not represent a random sample from the general population \citep{parkin2023facilitating,donnell2023study}.
People who are aware that they are HIV-positive may not participate in screening, resulting in an underrepresentation of infected individuals in the sample, and consequently, an underestimation of incidence.
One potential solution is to exclude individuals that have had an HIV test within the time period during which an infection would be considered ``recent'', often deemed to be one or two years preceding the screening time (the threshold for what is considered ``recent'' is based on the properties of the assay used for recency testing). 
However, this strategy is impractical, as it excludes a significant number of individuals, making trial recruitment challenging.
The recent PURPOSE 1 study, a phase 3, double-blind, randomized trial designed to evaluate the efficacy of twice-yearly subcutaneous lenacapavir among adolescent girls and young women in South Africa and Uganda, addressed this issue by excluding participants who had undergone HIV testing within the past three months \citep{bekker2024twice}. 
The study's primary findings were remarkable: no HIV infections were reported among participants receiving lenacapavir, demonstrating its superior efficacy and marking a significant advancement in HIV prevention. 
However, further investigation is warranted to assess the testing-based exclusion criterion's effectiveness in correcting for bias and to guide the designs of future studies.
This will help guide the design of future studies in HIV prevention.

In this paper, we evaluate the impact of using an ``testing-based criterion'': only including individuals who have not undergone HIV testing in a pre-specified period in the sample for cross-sectional HIV incidence estimation. We focus on the cross-sectional incidence estimator in \cite{kassanjee2012new}. Importantly, we consider non-representative sampling as a result of HIV-positive and negative individuals attending screening at different rates, which is the motivation for using the testing-based criterion. In Section \ref{sec:methodology}, we develop a statistical framework that incorporates both this biased sampling and the ``testing-based criteria'' to understand the behavior of the cross-sectional incidence estimator. We propose an effective mean duration of infection (MDRI) to evaluate and correct the bias in incidence estimation. In Section \ref{sec:simulation study}, we conduct simulation studies to evaluate the performance of the incidence estimation in realistic settings. We illustrate how excluding individuals based on prior test results can be advantageous in specific settings and provide recommendations on choosing testing-based criteria for screening in practice.

\section{Methodology } \label{sec:methodology}
\begin{figure}
    \centering
    \includegraphics[width=0.9\linewidth]{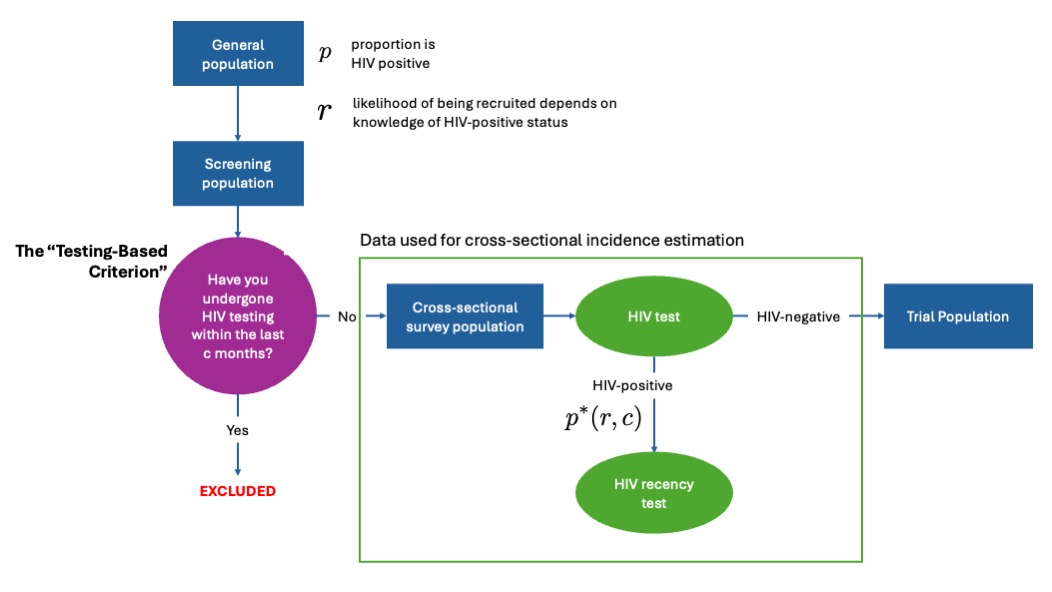}
    \caption{Cross-sectional incidence estimation and testing-based criteria in trial settings.}
    \label{fig:diagram}
\end{figure}

In this section, we introduce the statistical framework developed to incorporate testing-based criteria into HIV incidence estimation and outline how we evaluate estimation performance in a realistic setting. 
Throughout the following sections, we refer to Figure \ref{fig:diagram}, which illustrates the use of cross-sectional incidence estimators paired with an testing-based criterion in trials. After recruitment from the General Population (which may not be representative in terms of the composition of HIV-negative and HIV-positive individuals), the individuals in the Screening Population are assessed for eligibility based on inclusion and exclusion criteria. One of these criteria may be whether individuals have had a prior HIV test within a defined time period. We call this a \textit{testing-based criterion}. A testing-based criterion is proposed as a way to mitigate the issues that arise from non-representative sampling (will be introduced in Section \ref{sec:meth-history}) when constructing an incidence estimator based on the individuals in the Cross-Sectional Survey Population. Individuals in this population receive HIV tests, and if positive, an additional HIV recency test.
HIV-negative individuals may be enrolled in the trial. The information within the green rectangle in Figure \ref{fig:diagram} is used for cross-sectional HIV incidence estimation, which aims to estimate the incidence rate in the Trial Population in the absence of preventative therapies.

A testing-based exclusion uses information about the timing of prior HIV tests. Section \ref{sec:meth-history} describes two assumptions about an individuals' HIV testing history and what constitutes their ``most recent'' HIV test, which is used when evaluating testing-based exclusion. 
In Section \ref{sec:screening_pop} and Section \ref{sec:meth-cs}, we define the Screening Population and the Cross-Sectional Survey Population. For the latter, we introduce the data on ``current'' HIV status information and HIV recency test results, in reference to a fixed time point, $t_{cs}$.
In Section \ref{sec:hiv-incidence-est}, we define the HIV incidence estimator based on these data. In Section Section \ref{sec:effect_MDRI} we introduce a novel metric called the ``effective MDRI'', which provides insights into the estimator's bias under non-representative sampling and a given testing-based criterion.

\subsection{HIV Testing History Data}\label{sec:meth-history}
Suppose that there is a sequence of potential HIV tests administered at calendar times $\breve{\mathcal{T}} = \{\breve{T}_{1},\breve{T}_{2}, \dots, \breve{T}_{n} \}$, where $n$ is the number of HIV tests that an individual would take through out their lifetime. Let $I$ be the calendar HIV infection time for an individual.
It is assumed that the potential testing history is independent of the infection time, i.e., $\breve{\mathcal{T}}  \perp I$. Let $t_{cs}$ denote a fixed point in time, which can be thought of as the ``current'' time. When anchored at time $t_{cs}$, only some of the tests within the testing sequence $\breve{T}$ are observed. Assumptions \ref{assumption:id} and \ref{assumption:SWP} define specific scenarios and the corresponding set of tests in an individuals' history as they relate to both $t_{cs}$ and $I$. 

\begin{assumption} [Observation of Regular HIV Testing History] \label{assumption:id}
    The observed testing history $\mathcal{T}$ is $\mathcal{T}^{ID} := \{t: t\in \breve{\mathcal{T}}, t \le t_{cs}\}$ {such that $\mathcal{T} \perp I$}.
\end{assumption}
\begin{assumption}[Observation of Stop When Positive (SWP) HIV testing history] \label{assumption:SWP}
    The observed testing history $\mathcal{T}$ is $\mathcal{T}^{SWP} := \{t: t\in \breve{\mathcal{T}}, t\le I, t \le t_{cs}\} \cup _{\min} \{t\in \breve{\mathcal{T}}, t\ge I, t \le t_{cs}\}$.
\end{assumption}
Under Assumption 1, we observe all potential tests for an individual up until $t_{cs}$. As a result, the observed testing sequence is independent of their infection time.
Assumption 2 is a more practical assumption, stating that individuals will not take any more HIV tests once they have had an initial positive result. 
Under Assumption 2, the observed sequence of HIV testing times is no longer independent of the infection time $I$. Assumption 1 and 2 are mutually exclusive.

\begin{definition}[Time Since Most Recent HIV Test] \label{def:MR}
The time since the most recent HIV test for individual $i$ can be defined as the length of time between most recent HIV test time and the cross-sectional survey time $t_{cs}$, i.e.,
\begin{align*}
    T = t_{cs} - _{max}\{t: t\in \mathcal{T} \} \ge 0.
\end{align*}
Let $T^{\text{ID}}$ and $T^{\text{SWP}}$ be the length of time since the most recent HIV test under Assumption \ref{assumption:id} and Assumption \ref{assumption:SWP}, respectively.
\end{definition}
Under Assumption \ref{assumption:id}, \(T^{\text{ID}} \perp I\) because \(\mathcal{T}^{ID} \perp I\). 
However, under Assumption \ref{assumption:SWP}, $T^{SWP} \not\perp I$  (as $\mathcal{T}^{SWP}$ and $I$ are intrinsically related). Let $D = \Ind{}{ I\le t_{cs}}$ be the indicator of HIV status, where $D = 1$ indicates HIV positive and $D = 0$ indicates HIV negative at time $t_{cs}$. If $D = 1$, then we also define $U = t_{cs} - I$ as the infection duration by the time $t_{cs}$.


\subsection{Trial Screening Population} \label{sec:screening_pop}

We now consider a population of individuals that is being screened for a trial, which we call the \textit{Screening Population}. Without loss of generality, consider the screening time to be $t_{cs}$ for everyone. It is possible that the probability of attending screening may differ depending on whether an individual is aware of their HIV status at time $t_{cs}$. Let $Q$ be an indicator that is 1 if the individual  attends screening and 0 otherwise, and $\Delta = \Ind{}{U \ge T}$ be an indicator that an HIV-positive individual is aware that they have HIV.

\begin{definition}[Selective Attendance Ratio]\label{def:r}

Define $q_1$ and $q_0$ to be the probability of attending the screening for individuals whose most recent HIV test result was positive and negative, respectively, i.e., $q_1 = \Pr(Q = 1 | \Delta=1)$ and $q_0 = \Pr(Q = 1 | \Delta=0)$. We denote $r = q_1/q_0 \in [0,1]$ as the ratio of these probabilities and refer to it as the ``Selective Attendance Ratio''.
\end{definition}
The selective attendance ratio is defined as the ratio of the probability of individuals with an HIV-positive test result attending screening to the probability of individuals with an HIV-negative test result attending screening. We assume that $r \leq 1$, meaning that individuals who are HIV-positive do not have a higher propensity for attending screening than those who are HIV-negative.
When $r = 1$, knowledge of HIV status does not influence whether or not individuals attend screening. 
If $r<1$, as suggested by previous trials, a testing-based criterion (Definition \ref{definition:exclusion}) may be applied to align the trial screening population more closely with the general population in terms of HIV status.
\begin{definition}[Testing-Based Criterion] \label{definition:exclusion}
    Individuals who have taken a test within $(t_{cs}-c,t_{cs})$ are to be excluded from the screening population, where $c\ge 0$ is a pre-specified constant.
\end{definition}
Someone "meets" the testing-based criterion if they have not taken a test within the last c time period. For example, if $c$ is 6 months, individuals who have taken a test within the last 6 months are not eligible for the trial.
Let $C$ indicate whether the individual would be excluded based on the defined testing-based exclusion, i.e., $C = \Ind{}{T > c}$. In our example, $C = 1$ would indicate an individual having not taken a test within the last 6 months.

\subsection{Cross-Sectional Survey Population}\label{sec:meth-cs}
Consider a cross-sectional survey of $N$ independent individuals. Without loss of generality, assume that the cross-sectional sample is taken at time $t_{cs}$. For individual $i = 1, 2, \dots, N$, let $\tilde T_i$, $\tilde U_i$ represent the observed time since the last HIV test and the HIV infection duration, respectively.
Here, the tilde notation distinguishes these variables for individuals in the cross-sectional sample, as their distribution no longer matches that of the general population, as defined in Sections \ref{sec:meth-history} and \ref{sec:screening_pop}. 
Specifically, the joint distribution of $(\tilde T_i, \tilde U_i)$ is the same as that of $(T, U)$ conditional on $Q = 1$ and $C = 1$.
Let $\tilde D_i$ denote the observed infection status for  individual $i$, such that the total number of HIV-positive in the cross-sectional survey is given by 
$N_{\text{pos}} = \sum_{i=1}^N \tilde D_i$. Define
$N_{\text{neg}} = N - N_{\text{pos}}$,
as the number of HIV-negative individuals in the cross-sectional survey.

\subsection{Incidence Estimator of HIV Cross-sectional Survey Based on Recency Test}
\label{sec:hiv-incidence-est}
HIV recency tests are administrated to individuals who are HIV-positive in the cross-sectional survey population. Let $\tilde R_i$ indicate that the recency testing algorithm classifies the individual as being ``recently infected" and $\tilde R_i = 0$ otherwise, with the definition of ``recent infection'' below.
\begin{definition}[Recency Infection] \label{assumption:recent}
For HIV-positive individuals (with $\tilde D_i = 1$), they are likely to be identified as ``recently infected" by a recency test  at the survey if their infection time is within the last $T^*$ time period, i.e., $\tilde U_i\le T^*$. 
\end{definition}

 The cutoff $T^*$ defines ``recently infected" and non ``recently infected" subjects. Based on the cutoff time, two important characteristics of recency test can be defined. MDRI ($\Omega_{T^*}$) is the mean duration of infection among individuals who are classified as ``recently infected" by recency tests and FRR ($\hat \beta_{T^*}$) is the probability of given a HIV-positive individual whose infection duration is over $T^*$ ($\tilde U_i > T^*$), yet being classified as ``recently infected" through recency test. Define $N_{rec} = \sum_{i=1}^N \tilde R_i$ is the number of individuals who are detected to be ``recently infected" at the time of survey.

A widely used incidence estimation approach in cross-sectional survey is the estimator proposed by \cite{kassanjee2012new}. This estimator can be derived using HIV infection and recency test results among recruited subjects under certain assumptions \citep{gao2022statistical}. With a sample of $N$ individuals and the data obtained from the sample, the estimator can be expressed as:\\
\begin{align} \label{lambda}
    \Tilde{\lambda} = \frac{N_{rec} - N_{pos} \hat{\beta}_{T^*}}{N_{neg} (\hat{\Omega}_{T^*} - \hat{\beta}_{T^*} T^*)},
\end{align}
where the estimates of $\hat{\Omega}_{T^*}$ and $\hat{\beta}_{T^*}$ are obtained from the external data.
The estimator in \eqref{lambda} using data from the cross-sectional survey population at can be expressed using our notation as:
\begin{align*}
    \Tilde{\lambda} = \frac{\sum_{i=1}^N \tilde D_i \tilde R_i- \hat{\beta}_{T^*} \sum_{i=1}^N \tilde D_i }{\hat{\Omega}_{T^*} \sum_{i=1}^N (1-\tilde D_i)-\hat{\beta}_{T^*} T^*}.
\end{align*}

\subsection{Effective MDRI: Statistical Insights on Estimation Bias} \label{sec:effect_MDRI}
According to its definition, MDRI can be expressed as:
\begin{align*}
    \Omega_{T^*} = \int_{0}^{T^*} \phi(u)du,
\end{align*}
where $\phi(u):= \Pr(\tilde R_i=1|\tilde U_i=u, \tilde D_i=1)$ is the duration-specific test-recent probability, indicating the probability of a randomly selected person being classified as ``recently infected" given his specific time of HIV infection. The probability $\phi(u)$ only depends on the infection duration and is independent of the screening time $t_{cs}$.
We propose a novel quantity which we call \textit{effective MDRI}. Effective MDRI can help quantify and explain the bias in \eqref{lambda} that would be expected under different selective attendance ratios and test-based criterions. 
We consider the simple case when FRR is zero ($\beta_{T^*}=0$) to make the derivation of the effective MDRI 
straightforward.
We evaluate the property of the estimators with non-zero FRR in the sensitivity analysis described in section \ref{sec:sensitivity}. 
\begin{definition}[effective MDRI] \label{def_eff_MDRI}
When the FRR $\beta_{T^*} = 0$, we define the effective MDRI $\Omega_{T^*,\text{eff}}$ as
\begin{align}
\Omega_{T^*,\text{eff}}
 = \frac{\int_0^{T^*} \phi(u) \{ r \Pr(T \le u, T > c | U = u,D =1) + \Pr(T > u, T > c | U = u,D =1)\} du}{\Pr(T > c | D = 0)} .\label{effective_MDRI}
\end{align}
\end{definition}
The concept of effective MDRI is based on the idea that in the original incidence estimator $\tilde \lambda$, the MDRI estimate is obtained from external data with similar population characteristics. Effective MDRI is proposed to address misalignment between the screening population and the target population. The effective MDRI reflects the characteristics of the screening population, and thus could ``de-bias'' the incidence estimator when there is non-representative sampling. 
The incidence estimator using effective MDRI with zero FRR can be expressed as
\begin{align*}
   \hat \lambda = \frac{\sum_{i=1}^N \tilde D_i \tilde R_i}{\hat{\Omega}_{T,\text{eff}} \sum_{i=1}^N (1- \tilde D_i)} 
\end{align*}
Hence, comparing $\Omega_{T,\text{eff}}$ and $\Omega_{T^*}$ can help understand the bias of incidence estimation. An alternative way to express  effective MDRI is:
\begin{align}
    \Omega_{T^*,\text{eff}} &= \frac{\int_0^{T^*} \phi(u)  \Pr( T > c |U = u,  D =1) du}{\Pr(T > c | D = 0)} - (1-r) \frac{\int_0^{T^*} \phi(u) \Pr(T \le u, T > c |U = u,  D =1)du}{\Pr(T > c | D = 0)}.\label{effective_MDRI2}
\end{align}
Under Assumption \ref{assumption:id}, $  T^{\text{ID}} \perp  U$, such that the first term in the expression (\ref{effective_MDRI2}) can be reduced to $\Omega_{T^*}$ (See Appendix \ref{appendix: effMDRI}).
Therefore, when there is no selective attendance ($r=1$), the HIV incidence estimator is unbiased for any $c\ge0$.
When $r<1$, $\Omega_{T^*,\text{eff}} \le \Omega_{T^*}$, which suggests  that HIV incidence will be underestimated since the second expression in (\ref{effective_MDRI2}) is always non-negative.

Under Assumption 2, when the parameter \(r = 1\) and \(c < T^*\), it is likely that the estimator \(\tilde \lambda\) overestimates the true incidence \(\lambda\) since $\Omega_{T^*, \text{eff}} \ge \Omega_{T^*}$ in this case (See Appendix \ref{appendix: effMDRI} for the detail).
When $c\ge T^*$, once again $\Omega_{T^*, \text{eff}} =\Omega_{T^*}$, leading to unbiased estimation of incidence. As the degree of selective attendance increases (\(r\) decreases from 1 to 0), the effective MDRI is non-increasing. This change impacts the bias in \(\tilde{\lambda}\) in the following ways:
\begin{enumerate}
    \item \textbf{Case \(c \ge T^*\)}: In this scenario, the non-increasing effective MDRI indicates that $\tilde{\lambda}$ will likely underestimate \(\lambda\).
    \item \textbf{Case \(c < T^*\)}: When \(r = 1\), $\tilde{\lambda}$ tends to overestimate \(\lambda\). However, as \(r\) decreases, the comparison between \(\tilde{\lambda}\) and the true \(\lambda\) becomes more complex. It is plausible that \(\tilde{\lambda}\) may be unbiased for certain values of \(r < 1\).
\end{enumerate}
This statistical result provides insight into the behavior of the estimator under different scenarios and highlights the importance of understanding the selective attendance ratio \(r\) and choosing a test-based criterion \(c\) to minimize bias.
\section{Simulation Study} \label{sec:simulation study}
We perform a simulation study to assess the performance of the estimators when the HIV testing histories follow different assumptions (Assumption \ref{assumption:id} and \ref{assumption:SWP}) across a range of scenarios. Section \ref{simulation_process} outlines our the simulation procedure. Section 3.2 presents our primary simulation resultsSection 3.3 presents simulation results from three sets of sensitivity analysis.

\subsection{Data Simulation Procedure }  \label{simulation_process}
To generate the cross-sectional survey data at time $t_{cs}$, we follow the simulation procedure of \cite{gao2022statistical}, which we briefly describe here. We generate a fixed sample size of $N=5,000$, constant prevalence of $29\%$, and a constant incidence rate $\lambda = 0.032$ per person-year (see details of Assumption \ref{assumption:const prev incid} in Appendix \ref{appendix: effMDRI}). We use $\phi(u) = \{1 - F_{\mathrm{Gamma}}(u; \alpha = 0.352, \beta = 1.273)\}\Ind{}{u \le T^*}$ to model recency tests with a relatively short mean window period (MDRI) of 98 days and zero FRR (see Figure 1 in \cite{gao2022statistical}). This test-recent function mimics the properties of a recency test, with a higher probability of testing "recently infected" for shorter infection durations, gradually decreasing as the infection duration approaches $T^* = 2$ years, where the test-recent rate probability reaches $1.4\%$ at $t = T^*$. For $t > T^*$, the test-recent rate probability is truncated to zero to simulate the setting with zero FRR ($\beta^* = 0$).
The sensitivity analysis with a non-zero FRR is given in Section \ref{sec:sensitivity-frr}.
The procedure to generate estimates $\hat{\Omega}_{T^*}$ is described in Section 3.3 in \cite{gao2022statistical}.

We assume that the testing history $\breve{\mathcal{T}} = \{\breve{T}_{1}, \breve{T}_{2}, \breve{T}_{3} \dots, \breve{T}_{n}\}$ follows the homogeneous Poisson process with the inter-test time $X_{j} = \breve{T}_{j} - \breve{T}_{j-1}$ follows an exponential distribution with rate $\theta$ for $ j =2,3,\dots n $ starting with $X_{1} = \breve{T}_{1}$. 
We set $\theta = 0.4,1,1.5$, or $2$ evaluate a range of testing frequencies, spanning from an average of once every 2.5 years to twice per year.
Based on the sequence of testing history, we generate the most recent test time in Definition \ref{def:MR} under Assumption \ref{assumption:id} ($  T^{\text{ID}}$) or Assumption \ref{assumption:SWP} ($ T^{\text{SWP}}$) (see details in Appendix \ref{Appendix:testing history}). We evaluate the accuracy and variability of the estimator $\Tilde{\lambda}$ when the selective attendance ratio $r = 0, 0.3, 0.6$, or $1$ and the effect of the testing-based criterion $c =0.25,1,1.5$, or $2$.

\subsection{Simulation Results} \label{sec:sim_result}
\begin{figure}[!h]
\centering
    \begin{subfigure}[b]{0.9\textwidth}
        \centering
        \includegraphics[width=\textwidth]{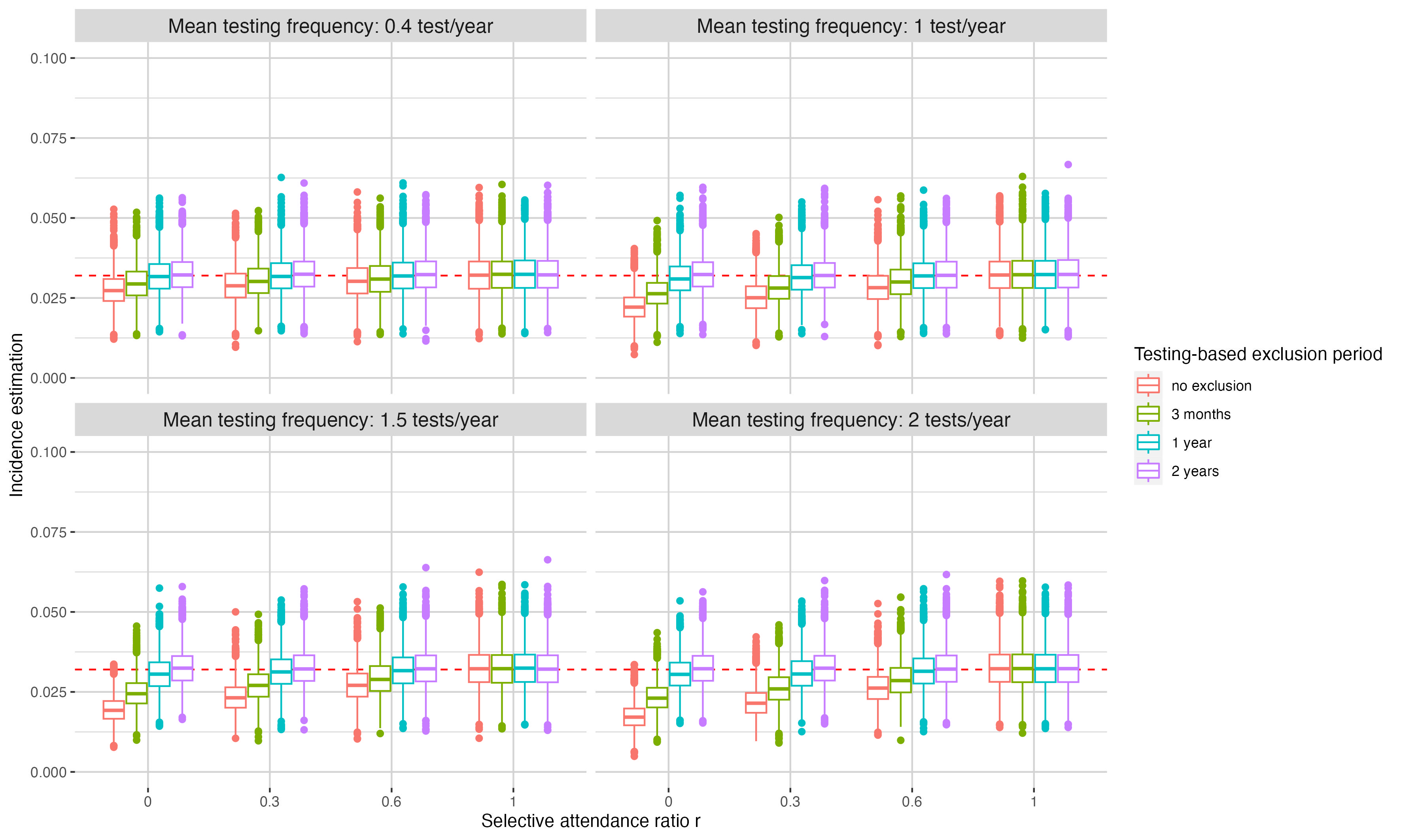}
        \caption{}
        \label{fig:box_A1}
    \end{subfigure}
    \hfill
    \begin{subfigure}[b]{0.9\textwidth}
    \centering
    \includegraphics[width=\textwidth]{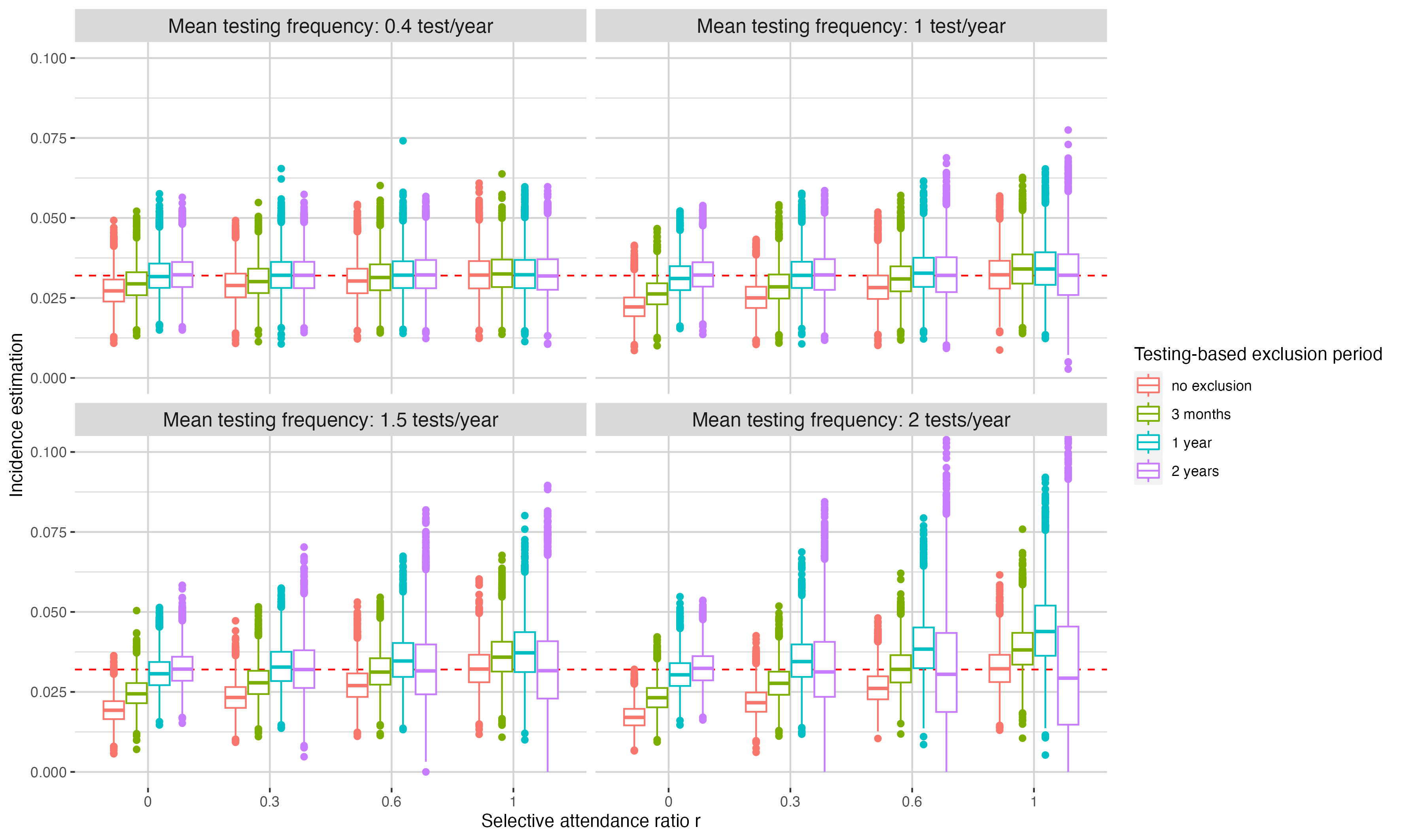}
    \caption{}
    \label{fig:box_A2}
    \end{subfigure}
    \caption{Performance of incidence estimation under (a) regular testing (Assumption \ref{assumption:id}) and (b) SWP testing (Assumption \ref{assumption:SWP}). Each subplot shows results for different mean testing frequencies $\theta$, with the x-axis representing the selective attendance ratio $r$ and different colors indicating various testing-based exclusion periods $c$. The true incidence rate is marked by the red line at 0.032.}
    \label{fig:box}
\end{figure}

\begin{figure}[h!]
    \centering
    \includegraphics[width=\linewidth]{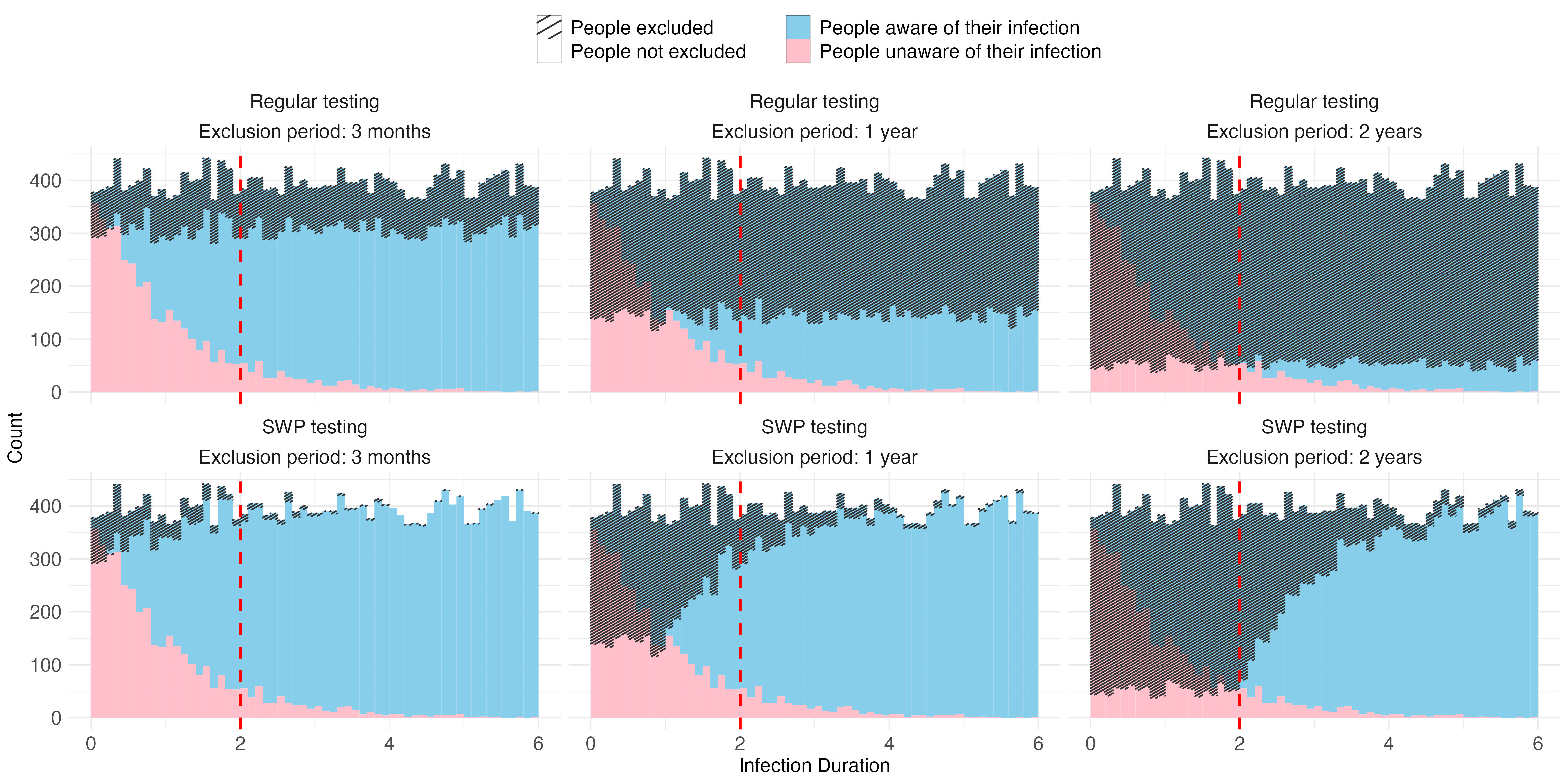}
    \caption{Composition of the HIV-infected population, broken down by their knowledge of their HIV status (color), their infection duration (x-axis), and whether they are excluded from the recency testing sample based on an testing-based criteria (pattern). The pink area represents the distribution of individuals aware of their HIV infections ($f_0(u,c)$), while the blue area represents the distribution of individuals unaware of their HIV infections ($f_1(u,c)$) under Assumption 1 and Assumption 2 with $\theta=1, q_0 = q_1 = 1$, using a sample size of 50,000 infected individuals. The influence of the testing-based criterion on $f_0(u,c)$ and $f_1(u,c)$ is depicted under testing-based criterion with $c = 0.25$, $c = 1$, and $c=2$ respectively. The red dashed line is at the time of $T^*$.}
    \label{fig:6_plots_theta1}
\end{figure}

Figure \ref{fig:box} (a) and (b) present boxplots of the incidence estimators under the regular testing (Assumption \ref{assumption:id}) and the SWP testing (Assumption \ref{assumption:SWP}) assumptions, respectively.
Each boxplot is based on $5,000$ simulations.
In addition, we present the diagnostic plots in Figure \ref{fig:6_plots_theta1} to illustrate the bias and variability under different scenarios. These plots reflect changes in the infection duration distribution, based on a total of 50,000 infected individuals and assuming no selective attendance ($r = 1$). The hatched areas represent the proportion of individuals excluded from the survey due to testing-based criteria.
To analyze bias, let $f_0(u;c) = \Pr(U = u, T > u, T > c | D = 1)$ and $f_1(u;c) = \Pr(U = u, T \le u, T > c | D = 1)$, which correspond to the included infection duration distributions of individuals aware of their infection (pink, non-hatched) and unaware of their infection (blue,non-hatched), respectively. 
Using $f_0(u;c) + r f_1(u;c)$ for $u < T^*$ provides insights into the bias of the estimator. The effective MDRI can be expressed as:
\[
\Omega_{T^*,\text{eff}} = \frac{p}{\lambda (1-p) } \frac{\int_0^{T^*} \phi(u) \p{f_0(u;c) + r f_1(u;c)}
du}{\Pr(T > c | D = 0)}.
\]
If the distribution of $f_0(u;c) + r f_1(u;c)$ for $u < T^*$ deviates from the shape of the original distribution of $U$, then $\Omega_{T^*,\text{eff}}$ may differ from $\Omega_{T^*}$, resulting in estimation bias. 
When $r = 1$, the non-hatched areas in the histogram represent the overall shape of the infection duration distribution in the survey. For $r < 1$, the blue non-hatched area is further proportionally reduced, reflecting the impact of selective attendance.

We list the main takeaways from our simulations results. Unless otherwise specified, we discuss the results under the more realistic Assumption \ref{assumption:SWP}. However, we note some key differences with the results under Assumption \ref{assumption:id}. Points (1), (2), and (3) below emphasize the balance between excluding individuals to account for selective attendance to screening, and high variance that may result from doing so.
\begin{enumerate}
    \item \textbf{Testing-based exclusion is not recommended when knowledge of HIV status does not influence screening attendance} \\
When knowledge of HIV status does not influence screening attendance ($r = 1$), the estimator is unbiased when no one is excluded $(c = 0)$ or when all individuals with a ``recent'' HIV test are excluded $(c = T^*)$. 
When we exclude individuals with a prior HIV test in the past $c \in (0, T^*)$ period of time, the incidence is often overestimated.
To achieve unbiased estimation (i.e., choosing from $c = 0$ and $c = T^*$), $c = 0$ is the clearly better choice, because $c = T^*$ excludes more individuals during the screening process, increasing the time it takes in order to recruit a target sample size in the cross-sectional survey. 

\item \textbf{When knowledge of HIV status influences screening attendance, the incidence estimator is biased unless everyone with a recent HIV test is excluded.} \\
When selective attendance based on HIV infection awareness is present ($r < 1$), the value of the estimator decreases compared to the cases without selective attendance ($r=1$).
Notably, selecting $c = T^*$ always ensures unbiased estimation.
This can also be seen in Figure \ref{fig:6_plots_theta1}, where the distribution of infection duration retains a uniform shape after excluding all individuals with a prior test less than $T^*$ ago ($c =T^*)$.
However, when $c\in (0,T^*)$, the distribution of infection duration is no longer uniform and shows a greater probability at $u>c$ (see Lemma \ref{SWP_dist} in Appendix \ref{appendix: effMDRI}).
In such cases, the estimator's behavior becomes more complex, as it may exhibit overestimation, unbiasedness, or underestimation depending on the parameter settings. Consequently, if the degree of selective attendance based on infection awareness is known, applying appropriate testing-based criteria on prior HIV testing history could help mitigate bias. Importantly, if no tesing-based exclusions are made ($c = 0$), the incidence estimator is biased, underscoring the importance of implementing the testing-based criterion on prior HIV testing history.

\item \textbf{In populations with very frequent testing, it may be difficult to find individuals who meet the test-based criterion.} 

As shown in Table \ref{table:population}, when frequency of HIV testing in the population is relatively low and the testing-based exclusion period is short (e.g., excluding individuals that took a test in the last 3 months), the number of individuals that need to be screened to achieve a cross-sectional survey of size $N = 5000$ remains manageable. However, as the testing-based criterion excludes more individuals, a large number need to be screened in order to achieve the same cross-sectional survey size. The testing-based criteria excludes more individuals when the testing frequency in the population is higher (increasing $\theta$) and/or and the length of the testing-based exclusion window ($c$) increases. For example, 
when individuals who have had a test within the last $T^*$ time period, the incidence estimator is unbiased under selective attendance. However, reaching a cross-sectional survey size of $N= 5,000$ may require screening at least $15,300$ individuals, a significant logistical effort.

\item \textbf{There is a bias-variance trade-off when choosing the testing-based exclusion window for prior HIV testing.}

Even beyond the effort required for screening to reach a desired sample size, certain settings may result in the incidence estimator having a prohibitively large variance, as shown in Figure 1(b) (e.g., when $c = T^*$).
This can be explained by the analytical form of the variance of $\log\tilde{\lambda}$, given by
\begin{equation}
    \text{Var} \left(\log(\tilde{\lambda})\right) = \frac{1}{N} \left( \frac{1}{p_r p^*} + \frac{1}{1-p^*} \right)\label{eq:var_lambda}
\end{equation}
where $p^*$ is the HIV prevalence in the cross-sectional survey population, and $p_r$ is the probability that an HIV-infected individual in that population is classified as recently infected by the recency testing algorithm \citep{gao2021sample}. Since $p_r p^* < p^*$ in all the scenarios, the first term in equation \eqref{eq:var_lambda} dominates the overall variability. As a result, the variance is primarily influenced by the fraction of HIV-positive individuals in the survey population who have been infected \textit{recently}. 

Although a testing-based criterion with $c = T^*$ results in unbiased estimation, it systematically reduces the proportion of recent HIV infections in the sample. 
Consider the scenario when $r = 1$ (there is no selective attendance), and $c = T^*$. Higher testing frequency in the population results in higher $p^*$ in the cross-sectional sample: HIV-positive individuals stop testing once they have a positive result, meaning their distribution of most recent tests is longer ago than the HIV-negative population, so they are less likely to be excluded by the testing-based criterion. At the same time, $p_r$ is very small, since the only HIV positive individuals who remain in the survey after the testing-based exclusion is applied either have very longstanding infections, or have been recently infected but have not yet taken an HIV test. Since $p_r \ll p^*$ in this scenario, changes in $p_r$ have a big impact on the variability of the incidence estimator (see the impact of changing $\theta$ on the population distribution in Appendix \ref{Appendix:disgnostic plots}).
Therefore, while the estimation remains unbiased in these scenarios, the variability becomes unacceptably high, rendering the incidence estimation unreliable.  These conclusions also hold when $\theta$ is fixed, but the testing-based exclusion period $c$ increases. This highlights an important bias-variance trade-off for choosing a suitable testing-based criterion (see the impact of changing $c$ on the population distribution in Figure \ref{fig:6_plots_theta1}).

When there is selective attendance ($r < 1$), the picture is different. The first term in \eqref{eq:var_lambda} becomes less influenced by $p_r$, as $p^*$ decreases due to fewer HIV-positive individuals attending screening and becomes closer in magnitude to $p_r$ (decreasing $r$). Consequently, the impact of $\theta$ and $c$ on the variance is attenuated. 
In the extreme case where HIV-positive individuals aware of their infection do not attend screening ($r = 0$), the variance remains relatively constant across different testing frequencies and testing-based exclusion periods.

\item \textbf{Using effective MDRI mitigates the bias, but requires distributional assumptions on an individual's testing history.} \\
If the distribution of individuals' HIV history is know, the effective MDRI can be calculated (see Appendix \ref{Appendix: effMDRI simulation} for details).
Since the effective MDRI accounts for the characteristics of the screened population,  replacing the MDRI with the effective MDRI in the incidence estimator ensures unbiased estimation.
However, it is important to note that this adjustment does not address the variability introduced to the estimator when the testing-based exclusion is applied.

\item \textbf{When HIV regular testing is observed (Assumption \ref{assumption:id}), the incidence estimator is more robust to a variety of testing-based exclusion strategies.} \\
Under the less realistic assumption of regular HIV testing history (Assumption \ref{assumption:id}), the estimator remains unbiased for any testing-based exclusion window ($c$) and testing frequency ($\theta$), provided there is no selective attendance based on individuals' awareness of their HIV status ($r = 1$). This difference in incidence performance between SWP testing (Assumption \ref{assumption:SWP}) and regular testing (Assumption \ref{assumption:id}) is illustrated in Figure \ref{fig:6_plots_theta1}. As shown in the figure, under Assumption \ref{assumption:id} and $r=1$, the distribution of infection duration within $[0,T^*]$ for infected individuals ains the same shape after applying the testing-based exclusion, indicating the absence of selection bias in this scenario (see details in Appendix \ref{appendix: effMDRI}). When there is selective attendance ($r < 1$), the shape of the distribution changes, as the blue non-hatched area proportionally decreases compared to the case when $r = 1$, leading to an underestimation of $\lambda$.
\end{enumerate}

\begin{table}[h!]
\centering
\scriptsize

\caption{People in the HIV cross-sectional survey under testing-based exclusion criteria when observing SWP testing. The bias are calculated analytically through effective MDRI and ``$\checkmark$" means the estimator is unbiased.}
\begin{tabular}{ccccc}
\toprule
Testing-based exclusion period & Mean testing frequency & Selective attendance ratio & bias ($\times 10^{-3}$) & Required screening people \\
\midrule
No exclusion & 1 test/year & $0$ & $-9.95$ & $5,000$ \\ 
 & 1 test/year & $60\%$ & $-3.98$ &  $5,000$  \\
  & 1 test/year & $100\%$ & $\checkmark$ &  $5,000$ \\
  & 2 tests/year & $0$ & $-15.03$ &  $5,000$ \\
  & 2 tests/year & $60\%$ & $-6.01 $ & $5,000$ \\
  & 2 tests/year & $100\%$ & $\checkmark$ & $5,000$ \\
  \hline
3 months & 1 test/year &  $0$ & $-5.93$  & $6,350$ \\
 & 1 test/year & $60\%$ & $-1.36$ &  $6,100$\\
 & 1 test/year & $100\%$ & $1.68$ &  $6,000$\\
 & 2 tests/year & $0$ & $-8.97$ &  $8,200$\\
 & 2 tests/year & $60\%$ & $-0.10$ &  $7,350$\\
 & 2 tests/year & $100\%$ & $5.82$ & $7,000$\\
  \hline
  2 years & 1 test/year & $0$ & $\checkmark$ &  $34,800$\\
  & 1 test/year & $60\%$ & $\checkmark$ &  $18,750$\\
  & 1 test/year & $100\%$ & $\checkmark$ &  $15,300$\\
  & 2 tests/year & $0$ & $\checkmark$ &  $25,685$ \\
  & 2 tests/year & $60\%$ & $\checkmark$ & $28,850$ \\
  & 2 tests/year & $100\%$ & $\checkmark$ &  $20,200$ \\
\bottomrule
\end{tabular}
\label{table:population}
\end{table}
\subsection{Sensitivity Analysis} \label{sec:sensitivity}
Here we conduct three additional simulation studies to evaluate the sensitivity of the incidence estimation under testing-based criteria to non-zero FRR ($\beta_{T^*}$) (section \ref{sec:sensitivity-frr}), HIV inter-test time following uniform distribution (section \ref{sec:sensitivity-uni}) and longer MDRI (section \ref{sec:sensitivity-phi1b}). The simulation results of the sensitivity studies can be found in Appendix \ref{appendix:sensitivity}.
\subsubsection{non-zero FRR $\beta_{T^*}$} \label{sec:sensitivity-frr}
To evaluate incidence estimation with a non-zero FRR, we simulate data with the recent probability function $\phi(u) = \{1 - F_{\text{Gamma}}(u; \alpha = 0.352, \beta = 1.273)\} \Ind{}{u \le T^*} + \beta_{T^*} \Ind{}{u > T^*}$ to mimic recency tests with the same MDRI but with a non-zero constant FRR. We set the FRR to $\beta_{T^*} \in \{0,0.5\%, 1\%, 2\%\}$ and present simulation results for selective attendance ratio $r \in \{0, 0.3, 0.6, 1\}$, mean testing frequencies $\theta \in \{0.4, 1\}$ and exclusion period $c\in\{0,T^*\} $ in Figure \ref{fig:box_frr}. 

When there is no selective attendance ($r = 1$), the estimators have a larger bias compared to the zero FRR case, consistent with the simulation results in \cite{gao2022statistical}.

With selective attendance ($r < 1$), the performance is similar to the zero FRR case, where biased estimation persists. Additionally, variability grows as FRR increases under both assumptions on observed HIV testing histories.

\subsubsection{HIV inter-test time following uniform distribution} \label{sec:sensitivity-uni}
In this section, we consider the sensitivity analysis where the individuals' HIV testing history follow a different distribution.
Specifically, the inter-test times follow a uniform distribution, i.e., $\breve T_{n} - \breve T_{n-1} \sim Uni[0,b]$, where $b = 3$ or $4$. 

Our simulation results of incidence estimation in Figure \ref{fig:box_uni} exhibit similar characteristics to those in Figure \ref{fig:box} in  Section \ref{sec:sim_result}. Testing-based exclusion  could help reduce bias due to selective attendance, but would induce variability. When testing frequency is lower, i.e., the inter-test window becomes wider, the variability of the incidence estimation are generally smaller. However, as testing frequency increases, the variability of the estimator also increase.
\subsubsection{Longer MDRI} \label{sec:sensitivity-phi1b}
We consider another simulation setting using a recency test with a longer MDRI (224 days).
Specifically, the recent probability function is given by $\phi(u) = \{1 - F_{\text{Gamma}}(u; \alpha = 0.681, \beta = 1.003)\} \Ind{}{u \le T^*} $ (see details in section 3.2, \cite{gao2022statistical}). Results in figure \ref{fig:box_1B} are consistent with the summary of the key properties in section \ref{sec:sim_result}. Compared to incidence estimation using recency tests with a shorter MDRI in previous simulations, the variability of estimation based on recency tests with a longer MDRI is relatively smaller overall. However, the bias becomes more pronounced when recency tests with a longer MDRI are applied with an testing-based exclusion period $c \in (0, T^*)$. This increased bias is influenced by the shape of the recent function $\phi(u)$ between $c$ and $T^*$, which determines the deviation from the true incidence rate.

\section{Discussion}

We investigated the performance of cross-sectional HIV incidence estimators based on recency testing under non-representative sampling and a criterion that excludes individuals based on their HIV testing history. 
We built a statistical framework that incorporates the testing-based criterion and introduced the effective MDRI metric, which mathematically quantifies bias.
To assess the performance of incidence estimators, we conducted extensive simulation studies under realistic settings.
Our findings indicate that excluding individuals based on prior HIV tests is unnecessary when knowledge of HIV status does not influence screening attendance (i.e., the sample is representative of the target population).
However, when HIV-positive individuals attend screening with a lower propensity than HIV-negative individuals, the incidence estimator becomes unreliable unless all individuals who have undergone recent HIV testing are excluded from the sample.
Additionally, we identified a trade-off between bias and variability associated with the use of an testing-based criterion: while excluding more individuals can reduce the bias caused by non-representative sampling, it also increases the variability of the incidence estimation. 
These results underscore the need for caution in applying testing-based criteria based on prior HIV testing. Understanding these dynamics is crucial for refining incidence estimation methods and enhancing the design of future HIV prevention trials. 

In this work, we carefully defined and distinguished two types of observed HIV testing histories: regular testing as defined in Assumption \ref{assumption:id}, and stop-when-positive (SWP) testing as outlined in Assumption \ref{assumption:SWP}.
We demonstrate that, while the testing-based criterion based on prior HIV testing can correct bias arising from unrepresentative sampling when Assumption \ref{assumption:id} holds, it fails under the more realistic SWP testing scenario. 
otably, the bias under SWP testing can be either positive or negative, depending on the specific characteristics of recency testing and testing frequency.

We introduced the selective attendance ratio to quantify the extent to which individuals attend screening based on their knowledge of their HIV status.
Such a parameter is generally unknown and we evaluated the performance of the estimators under varying values of this parameter in the simulation studies.  
In practice, this parameter (along with others related to testing frequency) may be partially inferred using available data from both screening programs and the general population. 
For instance, if HIV testing behavior observed in the screening data deviates significantly from that in the surveillance data for the same population (such as a notably longer interval since the most recent HIV test among screening participants compared to the general population) it could indicate that knowledge of HIV status affects screening attendance significantly.

Our findings indicate that applying an testing-based criterion to restrict the sample used for cross-sectional incidence estimation may not be the optimal solution to address non-representative sampling due to selective screening. Instead, the proposed effective MDRI metric offers a promising alternative to mitigate bias without encountering the bias-variability trade-off. 
However, applying this adjustment to the incidence estimator requires specific distributional assumptions about HIV testing history and accurate estimation of key parameters, including testing frequency ($\theta$) and the selective attendance ratio ($r$). Further work is needed to explore how to modify the incidence estimator under unrepresentative sampling conditions.

Our work has several limitations. First, while the statistical framework we developed relies on minimal distributional assumptions, the simulations depend on parametric assumptions about prior HIV testing histories, which may not fully capture real-world behaviors. Second, HIV testing behavior could be more complex than the assumed stop-when-positive testing. For instance, individuals might seek HIV testing shortly after unprotected sex, leading to a correlation of testing frequency with HIV infection. Future work should focus on refining incidence estimators to address unrepresentative sampling while also accounting for these more nuanced testing behaviors.

\newpage
\bibliography{ref.bib}{}

\begin{thebibliography}{}

\bibitem[Bannick et~al., 2024]{bannick2024enhanced}
Bannick, M., Donnell, D., Hayes, R., Laeyendecker, O., and Gao, F. (2024).
\newblock An enhanced cross-sectional hiv incidence estimator that incorporates prior hiv test results.
\newblock {\em Statistics in Medicine}.

\bibitem[Bekker et~al., 2024]{bekker2024twice}
Bekker, L.-G., Das, M., Abdool~Karim, Q., Ahmed, K., Batting, J., Brumskine, W., Gill, K., Harkoo, I., Jaggernath, M., Kigozi, G., et~al. (2024).
\newblock Twice-yearly lenacapavir or daily f/taf for hiv prevention in cisgender women.
\newblock {\em New England Journal of Medicine}, 391(13):1179--1192.

\bibitem[Busch et~al., 2010]{busch2010beyond}
Busch, M.~P., Pilcher, C.~D., Mastro, T.~D., Kaldor, J., Vercauteren, G., Rodriguez, W., Rousseau, C., Rehle, T.~M., Welte, A., Averill, M.~D., et~al. (2010).
\newblock Beyond detuning: 10 years of progress and new challenges in the development and application of assays for hiv incidence estimation.
\newblock {\em Aids}, 24(18):2763--2771.

\bibitem[Donnell et~al., 2023]{donnell2023study}
Donnell, D., Kansiime, S., Glidden, D.~V., Luedtke, A., Gilbert, P.~B., Gao, F., and Janes, H. (2023).
\newblock Study design approaches for future active-controlled hiv prevention trials.
\newblock {\em Statistical Communications in Infectious Diseases}, 15(1):20230002.

\bibitem[Gao and Bannick, 2022]{gao2022statistical}
Gao, F. and Bannick, M. (2022).
\newblock Statistical considerations for cross-sectional hiv incidence estimation based on recency test.
\newblock {\em Statistics in medicine}, 41(8):1446--1461.

\bibitem[Gao et~al., 2021]{gao2021sample}
Gao, F., Glidden, D.~V., Hughes, J.~P., and Donnell, D.~J. (2021).
\newblock Sample size calculation for active-arm trial with counterfactual incidence based on recency assay.
\newblock {\em Statistical Communications in Infectious Diseases}, 13(1):20200009.

\bibitem[Grant et~al., 2010]{grant2010preexposure}
Grant, R.~M., Lama, J.~R., Anderson, P.~L., McMahan, V., Liu, A.~Y., Vargas, L., Goicochea, P., Casap{\'\i}a, M., Guanira-Carranza, J.~V., Ramirez-Cardich, M.~E., et~al. (2010).
\newblock Preexposure chemoprophylaxis for hiv prevention in men who have sex with men.
\newblock {\em New England Journal of Medicine}, 363(27):2587--2599.

\bibitem[Kaplan and Brookmeyer, 1999]{kaplan1999snapshot}
Kaplan, E.~H. and Brookmeyer, R. (1999).
\newblock Snapshot estimators of recent hiv incidence rates.
\newblock {\em Operations Research}, 47(1):29--37.

\bibitem[Kassanjee et~al., 2012]{kassanjee2012new}
Kassanjee, R., McWalter, T.~A., B{\"a}rnighausen, T., and Welte, A. (2012).
\newblock A new general biomarker-based incidence estimator.
\newblock {\em Epidemiology}, 23(5):721--728.

\bibitem[Kim et~al., 2019]{kim2019tracking}
Kim, A.~A., Behel, S., Northbrook, S., and Parekh, B.~S. (2019).
\newblock Tracking with recency assays to control the epidemic: real-time hiv surveillance and public health response.
\newblock {\em Aids}, 33(9):1527--1529.

\bibitem[Landovitz et~al., 2021]{landovitz2021cabotegravir}
Landovitz, R.~J., Donnell, D., Clement, M.~E., Hanscom, B., Cottle, L., Coelho, L., Cabello, R., Chariyalertsak, S., Dunne, E.~F., Frank, I., et~al. (2021).
\newblock Cabotegravir for hiv prevention in cisgender men and transgender women.
\newblock {\em New England Journal of Medicine}, 385(7):595--608.

\bibitem[Lau et~al., 2022]{lau2022systematic}
Lau, J. K.-O., Murdock, N., Murray, J., Justman, J., Parkin, N., and Miller, V. (2022).
\newblock A systematic review of limiting antigen avidity enzyme immunoassay for detection of recent hiv-1 infection to expand supported applications.
\newblock {\em Journal of Virus Eradication}, 8(3):100085.

\bibitem[McDougal et~al., 2006]{mcdougal2006comparison}
McDougal, J.~S., Parekh, B.~S., Peterson, M.~L., Branson, B.~M., Dobbs, T., Ackers, M., and Gurwith, M. (2006).
\newblock Comparison of hiv type 1 incidence observed during longitudinal follow-up with incidence estimated by cross-sectional analysis using the bed capture enzyme immunoassay.
\newblock {\em AIDS Research \& Human Retroviruses}, 22(10):945--952.

\bibitem[Parkin et~al., 2023]{parkin2023facilitating}
Parkin, N., Gao, F., Grebe, E., Cutrell, A., Das, M., Donnell, D., Duerr, A., Glidden, D.~V., Hughes, J.~P., Murray, J., et~al. (2023).
\newblock Facilitating next-generation pre-exposure prophylaxis clinical trials using hiv recent infection assays: a consensus statement from the forum hiv prevention trial design project.
\newblock {\em Clinical Pharmacology \& Therapeutics}, 114(1):29--40.

\end{thebibliography}
\bibliographystyle{apalike}

\newpage
\appendix
\section*{Appendix}
\addcontentsline{toc}{section}{Appendix} 
\section{Derivation of Effective MDRI: Statistical Insights on Estimating Bias} \label{appendix: effMDRI}
\begin{assumption}[Constant prevalence and constant incidence]  \label{assumption:const prev incid}
For $u \in[0, \ _{max}(\tau, T^*)], \Pr(U = u | D = 1) = \Pr(U = 0 | D = 1)$, with $\tau >0$ be the bounded support for the time between $T$ and time of survey $t_{cs}$.
\end{assumption}
Assumption \ref{assumption:const prev incid} is required for the incidence estimator in \cite{kassanjee2012new}: both incidence and prevalence remain constant throughout the study period \cite{gao2022statistical}. 
To formally derive and compare the property of the HIV incidence estimator under different scenarios, we define the effective MDRI in Definition \ref{def_eff_MDRI}, which quantifies the estimation bias. 
The expression of effect MDRI can be derived as
\begin{align*}
\Omega_{T^*,\text{eff}}
&= \frac{E[\tilde D_i \tilde R_i]}{\lambda E[(1-\tilde D_i)]} \\
&= \frac{E[\tilde D_i]}{\lambda E[ 1-\tilde D_i]} \int_0^{T^*} \Pr(\tilde R_i = 1, \tilde U_i = u | \tilde D_i = 1) du \\
&= \frac{E[\tilde D_i]}{\lambda E[ 1-\tilde D_i]} \int_0^{T^*} \phi(u) \Pr(\tilde U_i = u | \tilde D_i = 1) du\\
&= \int_0^{T^*} \phi(u) \frac{\Pr(\tilde D_i=1,\tilde U_i = u)}{\lambda \Pr(\tilde D_i = 0)}du\\
&= \int_0^{T^*} \phi(u) \frac{\Pr(U = u,D=1, Q=1, C=1)}{\lambda \Pr(D = 0,Q=1,C=1)}du\\
&=\int_0^{T^*} \phi(u) \frac{p}{1-p}\frac{\Pr(U = u,Q=1, C=1|D=1)}{\lambda \Pr( Q=1,C=1|D=0)}du\\
&= \int_0^{T^*} \phi(u) \frac{p}{1-p}\frac{q_1\Pr(U = u,\Delta=1, C=1|D=1) + q_0 \Pr(U = u,\Delta=0, C=1|D=1)}{\lambda q_0 \Pr( \Delta=0,C=1|D=0)}du\\
&= \frac{p}{\lambda (1-p)}\frac{\int_0^{T^*} \phi(u) \{ r \Pr(U = u, T \le u, T > c | D =1) + \Pr(U = u, T > u, T > c | D =1)\} du}{\Pr(T > c | D = 0)} \\
\text{(Lemma \ref{dist_U})} \ &= \frac{\int_0^{T^*} \phi(u) \{ r \Pr(T \le u, T > c | U = u, D =1) + \Pr(T > u, T > c | U = u, D =1)\} du}{\Pr(T > c | D = 0)} \\
&= \frac{\int_0^{T^*} \phi(u)  \Pr(T > c | U=u,D_i =1) du}{\Pr(T > c | D = 0)} - (1-r) \frac{\int_0^{T^*} \phi(u) \Pr(T \le u, T > c | U = u,D =1)du}{\Pr(T > c | D = 0)},
\end{align*}
where p is the constant prevalence and $\lambda$ is the true incidence rate among the general population. 
Under Assumption \ref{assumption:id}, $T^{\text{ID}} \perp U$, such that the first expression in (\ref{effective_MDRI2}) can be reduced to
\begin{align*}
    \frac{\int_0^{T^*} \phi(u)  \Pr(T^{\text{ID}} > c | U=u, D =1) du}{\Pr(T^{\text{ID}} > c | D = 0)} =     \frac{\int_0^{T^*} \phi(u)  \Pr(T^{\text{ID}} > c | D =1) du}{\Pr(T^{\text{ID}} > c | D = 0)} =\Omega_{T^*}. 
\end{align*}
Therefore, when $r=1$, $\Omega_{T^*,\text{eff}} = \Omega_{T^*}$, indicating unbiased estimation for any $c\ge0$.
When $r<1$ and $c<T^*$, $\Omega_{T^*,\text{eff}} \le \Omega_{T^*}$, suggesting potential underestimation since the second expression in (\ref{effective_MDRI}) is always non-negative. When  $r<1$ and $c \ge T^*$, since $u\le T^*, \Pr(U = u, c<T\le u | D=1) = 0$. The second expression in (\ref{effective_MDRI}) therefore disappears and this leads to unbiased estimation.\\
\\
Under Assumption 2, and $c<T^*$, the first expression in (\ref{effective_MDRI}) can be written as:
\begin{align}
    \Omega_{T^*,\text{eff}} &=\int_0^{T^*} \phi(u) \frac{\Pr(T^{\text{SWP}}>c | U=u, D = 1) }{\Pr(T^{\text{SWP}} > c | D = 0)} du \nonumber \\
    &= \int_0^{c} \phi(u) \frac{\Pr(T^{\text{SWP}}>c | U=u, D = 1) }{\Pr(T^{\text{SWP}} > c | D = 0)} du + \int_c^{T^*} \phi(u) \frac{\Pr(T^{\text{SWP}}>c | U=u, D = 1) }{\Pr(T^{\text{SWP}} > c | D = 0)} du \nonumber\\ 
 \text{(Lemma \ref{SWP_dist})}   &\ge \left\{\int_0^{c} \phi(u) \frac{\Pr(T^{\text{ID}}>c| D=1)}{\Pr(T^{\text{ID}} > c | D = 0)} du  + \int_c^{T^*} \phi(u) \frac{\Pr(T^{\text{ID}}>c| D=1) }{\Pr(T^{\text{ID}} > c | D = 0)} du\right\}  \nonumber \\
 &= \Omega_{T^*}  \nonumber
\end{align}
Therefore, under Assumption 2, when the parameter \(r = 1\) and \(c < T^*\), it is likely that the estimator \(\tilde{\lambda}\) overestimates the true parameter \(\lambda\). This is due to the effective MDRI being positively biased.
As for $c\ge T^*$, the expression would be reduced to 
\begin{align*}
    \Omega_{T^*,\text{eff}} &= \int_0^{T^*} \phi(u) \frac{\Pr(T^{\text{ID}}>c| D=1) }{\Pr(T^{\text{ID}} > c | D = 0)} du = \Omega_{T^*},
\end{align*}
leading to unbiased estimation of MDRI.\\
\begin{lemma} \label{SWP_dist} For any $0\le u<c$, 
\[\Pr(T^{\text{SWP}}>c|U=u D=1) = \Pr(T^{\text{ID}}>c| D=1);
\]
for any $c<u<T^*$,
\[
\Pr(T^{\text{SWP}}>c| U=u, D=1) \ge \Pr(T^{\text{ID}}>c| D=1).
\]
 
\end{lemma}
{\it Proof.} Note that, 
\begin{align}
    T^{\text{SWP}} = \mathbbm{1}(U<T^{\text{ID}})T^{\text{ID}} + \mathbbm{1}(U\ge T^{\text{ID}}) T'. \label{eq: T_swp- T_id}
\end{align}
for some $T'$ that satisfies $U\ge  T'\ge T^{\text{ID}}$. 
That is, if an individual is test-negative at the most recent HIV test under Assumption 1, then the most recent test under Assumption 2 keeps the same; otherwise, it is likely that the most recent HIV test under Assumption 2 is further away from the cross-sectional time, however, it is always after the HIV infection time.
Therefore, for any $u,c>0$, we have
\[\Pr(T^{\text{SWP}}>c| U=u,D=1) =  \Pr(T^{\text{ID}}>c, T^{\text{ID}}>u| U=u, D=1) + \Pr(T' >c, T^{\text{ID}}\le u | U=u, D=1).\]

For any $u<c$, the second term is always zero, such that
\begin{align*}
    \Pr(T^{\text{SWP}}>c| U=u ,D=1) =&  \Pr(T^{\text{ID}}>c|U=u,D=1)=\Pr(T^{\text{ID}}>c| D=1),
\end{align*}
where the last equality holds due to the independence of $U$ and $T^{\text{ID}}$.
For any $c<u<T^*$,
\begin{align*}
    & \Pr(T^{\text{SWP}}>c|U=u, D=1) \\
    &= \Pr(T^{\text{ID}}>c| U=u, D=1) - \Pr( c<T^{\text{ID}} \le u|U=u, D=1) + \Pr(T' >c,T^{\text{ID}} \le u| U=u,D=1) \\
    &= \Pr(T^{\text{ID}}>c| U=u, D=1) + \Pr(T^{\text{ID}}\le c<T', T^{\text{ID} }\le u|U=u, D=1) \\
    &\ge \Pr(T^{\text{ID}}>c| U=u, D=1).
\end{align*}

\begin{lemma} \label{dist_U}
With constant infection rate under Assumption \ref{assumption:const prev incid}, we have $\Pr(U_i = u | D_i = 1) = \frac{\lambda (1-p)}{p}, \forall u>0$ as defined in \cite{gao2022statistical}.
\end{lemma}

\section{Simulation procedure of HIV testing history } \label{Appendix:testing history}
\subsection{Generating Duration of Most Recent Test}
\begin{assumption}[HIV Testing History Following Renewal Process] \label{assumption:test_hist}
The testing behaviours for individuals in the survey follow a renewal process. 
\end{assumption}
With Assumption \ref{assumption:test_hist}, the sequence of variables $\{  X_{1}=\breve T_{1},  X_{2}=\breve T_{2}-\breve T_{1}, \dots  X_{n} = \breve T_{n}- \breve T_{n-
1} \}$ representing inter-arrival time for two sequential HIV tests, are independent and identically distributed, and follow a specific distribution F, i.e., $ X_{1},  X_{2},\dots X_{n} \sim F$ with finite expectation $\mu < \infty$. 
Based on Assumption \ref{assumption:test_hist}, we can use the following limit aging theorem to generate $\breve T$ based on the distribution of inter-test time $\{X_{n}\}$:
\begin{lemma}[Renewal Limit Theorem for Aging Process] \label{lemma:renewal}
     For a sequence of interarrival time $X=\{X_1, X_2, \dots, X_n,\dots\}$ which are independent, identically distributed, nonnegative variables with common distribution function $F$ and finite expectation $E(X) = \mu <\infty$. Let $T = \{T_1,T_2,\dots, T_n, \dots\}$ with $T_n = \sum_{j=1}^n X_j$ be the arrival time sequence. For a certain time point $t>0$, let $T_{n(t)}$ be the most recent arrival time before t, i.e., $n(t) = _{max}\{n(t): T_{n(t)} < t < T_{n(t)+1}, n(t) \in \N \}$. Then 
     $\lim_{t \to \infty} \Pr (t- T_{n(t)}<x) = \lim_{t \to \infty} \Pr (T_{n(t)+1}-t <x) = \frac{1}{\mu} \int_0^x (1-F(y))dy, \ x\in [0,+\infty]$.
\end{lemma} 

When the inter-arrival time follows the exponential distribution with constant intensity $\theta>0$, i.e., $X_i \sim \text{Exponential} (\theta)$, the distribution of the time between the last test and current moment also follows the exponential distribution with intensity $\theta$: $t-T_{n(t)} \sim \text{Exponential} (\theta)$ since $\lim_{t\to \infty} \frac{\partial}{\partial x}\Pr(t-T_{n(t)} <x) = \frac{1}{\mu} (1-F(x)) = \theta e^{-\theta x}, x>0$. 

In sensitivity analysis, we let $X_j \sim \text{Uni}[a,b]$, with $\frac{1}{\mu} = \frac{2}{a+b}, \ 1-F(x) = \Ind{}{x<a} + \frac{b-x}{b-a} \Ind{}{a\le x\le b}$. Therefore, 
\begin{align*}
    \lim _{t\to \infty}\Pr(t-T_{n(t)} < x) 
    =\frac{2}{a+b} \{\int_0^x \Ind{}{y<a} + \frac{b-y}{b-a} \Ind{}{a\le y\le b} dy \} 
    = \frac{2x}{a+b}  \Ind{}{x<a} + \frac{-x^2 + 2bx - a^2}{b^2-a^2} \Ind{}{a\le x\le b}  + \Ind{}{x > b} 
\end{align*}
Then we can randomly generate the time between the calendar time of infection and the next HIV test time using inverse transform sampling. Let $e \sim \text{Uni}[0,1]$ and $e = \frac{2x}{a+b}  \Ind{}{x<a} + \frac{-x^2 + 2bx - a^2}{b^2-a^2} \Ind{}{a\le x\le b}  + \Ind{}{x > b} $, 
\begin{align*}
    x =  \frac{(a+b)e}{2}\Ind{}{0 \le e< \frac{2a}{a+b}} +(b-\sqrt{(b^2-a^2) (1-e)})\Ind{}{\frac{2a}{a+b} \le e \le 1}.
\end{align*}

To generate $T^{\text{ID}}$, we directly apply Lemma \ref{lemma:renewal}, setting the current time $t$ to $t_{cs}$. To generate $T^{\text{SWP}}$ while maintaining the relation between $T^{\text{ID}}$ and $T^{\text{SWP}}$ in (\ref{eq: T_swp- T_id}), we proceed as follows: if $T^{\text{ID}} \ge U$, set $T^{\text{SWP}} = T^{\text{ID}}$. If $T^{\text{ID}} < U$, generate additional tests starting from $T^{\text{ID}}$ until reaching the last test before $U$, which becomes the desired $T^{\text{SWP}}$.

\subsection{Simulating Population in HIV Cross-sectional Survey}
Let $s=\Pr(C = 1| Q = 1)= \Pr(Q = 1, C =1) / \Pr(Q = 1)$ be the probability of a coming individual getting included in the survey. 
First we derive $E[Q C]$,
\begin{align}
    E[QC] &= E[[E[Q | \Delta] E[C| \Delta]]] \nonumber \\
&= q_1E[\Delta] + q_0E[(1-\Delta) C]  \nonumber \\
&=q_1\Pr(U \ge T, T > c) + q_0 \Pr(U < T,T > c) \nonumber \\
&=q_1 \Pr(U \ge T, T >c |D = 1)\Pr(D=1) +q_0\p{\Pr(T > U, T> c, D = 1) + \Pr(T> c, D = 0) } \nonumber \\
&= q_1\int_0^{t^*}\Pr(U \ge T, T>c | U = u, D = 1) \Pr(U = u|D = 1) \Pr(D = 1) du \nonumber \\
    &\ + q_0 \p{\int_0^{t^*} \Pr(T> , T > c| U = u, D = 1) \Pr(U = u | D = 1) \Pr(D = 1) du + \Pr(T > c, D = 0) }\nonumber \\
    &= \lambda (1-p) \p{q_1\int_c^{t^*} \int_c^u \Pr(T = t | U = u, D =1) dtdu + q_0\int_0^{t^*} \int_{\max(u,c)}^\infty  \Pr(T = t|U = u, D = 1) du } \nonumber \\
    &\ + q_0\Pr(T > c, D = 0),\label{eq:exp_QiCi}
\end{align}
where $t^*$ is defined to assure equivalent incidence at a range of times closer to $t_{cs}$ (see details in the definition of $c_t$ in supplementary material, \cite{gao2022statistical}).

Then, if regular testing is observed, we have $T^{\text{ID}} \sim \text{Exponential}(\theta)$ and $T^{\text{ID}} \perp U$, (\ref{eq:exp_QiCi}) can be further derived as
\begin{align*}
& \lambda (1-p) \p{q_1\int_c^{t^*} \int_c^u \Pr(T^{\text{ID}} = t | U = u, D =1) dtdu 
 + q_0\int_0^{t^*} \int_{\max(u,c)}^\infty  \Pr(T^{\text{ID}} = t|U = u, D = 1) du }  \\
&\ + q_0\Pr(T^{\text{ID}} > c, D = 0) \\
&= \lambda (1-p) \p{q_1 \int_c^{t^*} \int_c^u \theta e^{-\theta t}dtdu + q_0 \int_0^{c} \int_{c}^\infty \theta e^{-\theta t} dtdu + q_0 \int_c^{t^*} \int_{u}^\infty \theta e^{-\theta t} dtdu + } + q_0(1-p)e^{-\theta c} \\
 &= \lambda (1-p) (q_1-q_0)(\frac{1}{\theta} e^{-\theta t^*} - \frac{1}{\theta} e^{-\theta c} - ce^{-\theta c}) + e^{-\theta c} \left\{q_1 p + q_0 (1-p)\right\}
\end{align*}
When SWP testing is observed, under Assumption \ref{assumption:SWP} and Lemma \ref{lemma:renewal}, $T^{\text{SWP}} $ has the piecewise conditional distribution
\begin{align}
    \Pr(T^{\text{SWP}} = t | U = u, D=1) &=\theta e^{-\theta(u-t)} \Ind{}{u\ge t} + \theta e^{-\theta t}\Ind{}{u<t},u\ge 0, t\ge 0; \label{eq:swp_dist1}\\
    \Pr(T^{\text{SWP}} = t |D=0) &= \theta e^{-\theta t}, t\ge 0. \label{eq:swp_dist2}
\end{align}
The expression in (\ref{eq:exp_QiCi}) under SWP testing can be further derived as
\begin{align*}
& \lambda (1-p) \p{q_1\int_c^{t^*} \int_c^u \Pr(T^{\text{SWP}} = t | U = u, D =1) dtdu + q_0\int_0^{t^*} \int_{\max(u,c)}^\infty  \Pr(T^{\text{SWP}} = t|U = u, D = 1) du } \\
&\ + q_0\Pr(T^{\text{SWP}} > c| D = 0) \Pr(D = 0) \\
  &=  \lambda (1-p) \p{q_1 \int_c^{t^*} \int_c^u \theta e^{-\theta (u-t)} dtdu + q_0 \p{\int_0^c \int_c^\infty \theta e^{-\theta t} dtdu + \int_c^{t^*} \int_u^\infty \theta e^{-\theta t} dtdu} }+ q_0 e^{-\theta c} (1-p)\\
  &= \lambda(1-p) \left\{ q_1 \p{ \frac{1}{\theta} e^{\theta c - \theta t^*} - c - \frac{1}{\theta}} +q_0 \p{ c e^{-\theta c} - \frac{1}{\theta} e^{-\theta t^*} + \frac{1}{\theta} e^{-\theta c}}\right\} + q_1p + q_0 e^{-\theta c}(1-p).
\end{align*}
The probability of an individual attending screening being expressed as
\[
\Pr(Q = 1) = \lambda (1-p)(q_1 - q_0)\frac{1}{\theta}(e^{-\theta t^*} - 1) + q_0(1-p) + q_1 p.
\]
Let $s^{\text{ID}}$ and $s^{\text{SWP}}$ be the probability of coming individuals getting included when observing regular testing or stop when positive testing. 
Then the probability of an attending individual being included under two different observed HIV testings can be expressed as:
\begin{align*}
    s^{\text{ID}} 
    &= \frac{\lambda (r-1)(\frac{1}{\theta} e^{-\theta t^*} - \frac{1}{\theta} e^{-\theta c} - ce^{-\theta c}) + e^{-\theta c} \left\{rp/(1-p) + 1\right\} }{ \lambda (r - 1)\frac{1}{\theta}(e^{-\theta t^*} - 1) + 1 + r p/(1-p)} ,\\
    s^{\text{SWP}} 
    &= \frac{\lambda \left\{ r \p{ \frac{1}{\theta} e^{\theta c - \theta t^*} - c - \frac{1}{\theta}} + \p{ c e^{-\theta c} - \frac{1}{\theta} e^{-\theta t^*} + \frac{1}{\theta} e^{-\theta c}}\right\} +  e^{-\theta c} + r p/(1-p)}{\lambda(r - 1)\frac{1}{\theta}(e^{-\theta t^*} - 1) + 1+ r p/(1-p)},
\end{align*}
respectively. Consequently, the required number of individuals to be screened in a survey with size $N$ under regular and SWP testing can be expressed as
$\frac{N}{s^{\text{ID}}}$ and $\frac{N}{s^{\text{SWP}}}$
respectively.

\section{Diagnostic plots under different HIV testing frequencies} \label{Appendix:disgnostic plots}
\begin{figure}[H]
    \centering
    \includegraphics[width=0.95\linewidth]{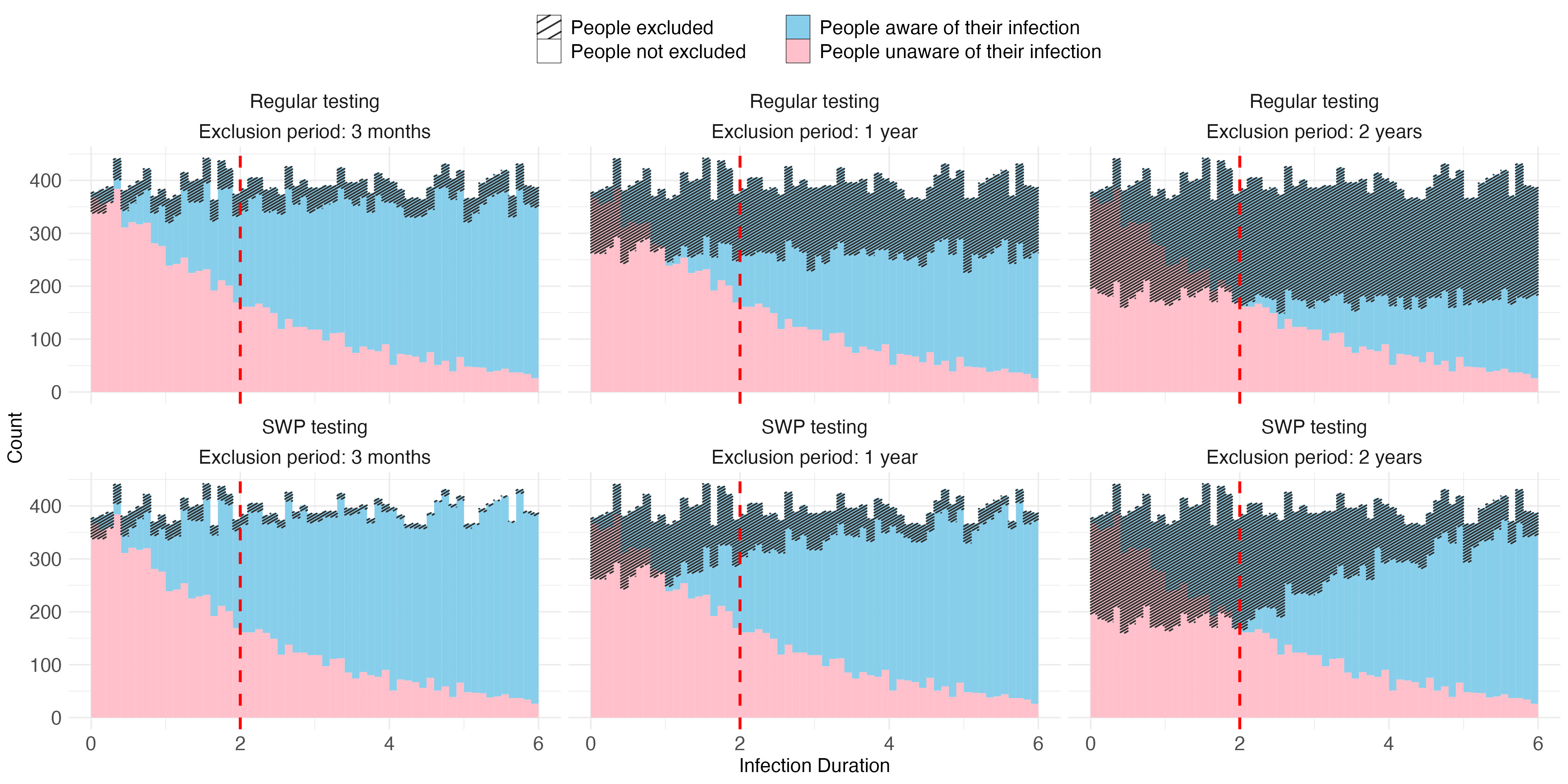}
    \caption{Composition of the HIV-infected population, broken down by their knowledge of their HIV status (color), their infection duration (x-axis), and whether they are excluded from the recency testing sample based on an testing-based criteria (pattern). The pink area represents the distribution of individuals aware of their HIV infections ($f_0(u,c)$), while the blue area represents the distribution of individuals unaware of their HIV infections ($f_1(u,c)$) when regular testing or SWP testing is observed with mean testing frequency of 0.4 test/year, using a sample size of 50,000 infected individuals. The influence of the testing-based criterion on $f_0(u,c)$ and $f_1(u,c)$ is depicted under testing-based criterion with $c = 0.25$, $c = 1$, and $c=2$ respectively. The red dashed line is at the time of $T^*$.}
    \label{fig:6_plots_theta025}
\end{figure}

\subsection{Diagnostic histogram of mean testing frequency of 0.4 test/year (Figure \ref{fig:6_plots_theta025})}
\begin{figure}[H]
    \centering
    \includegraphics[width=0.95\linewidth]{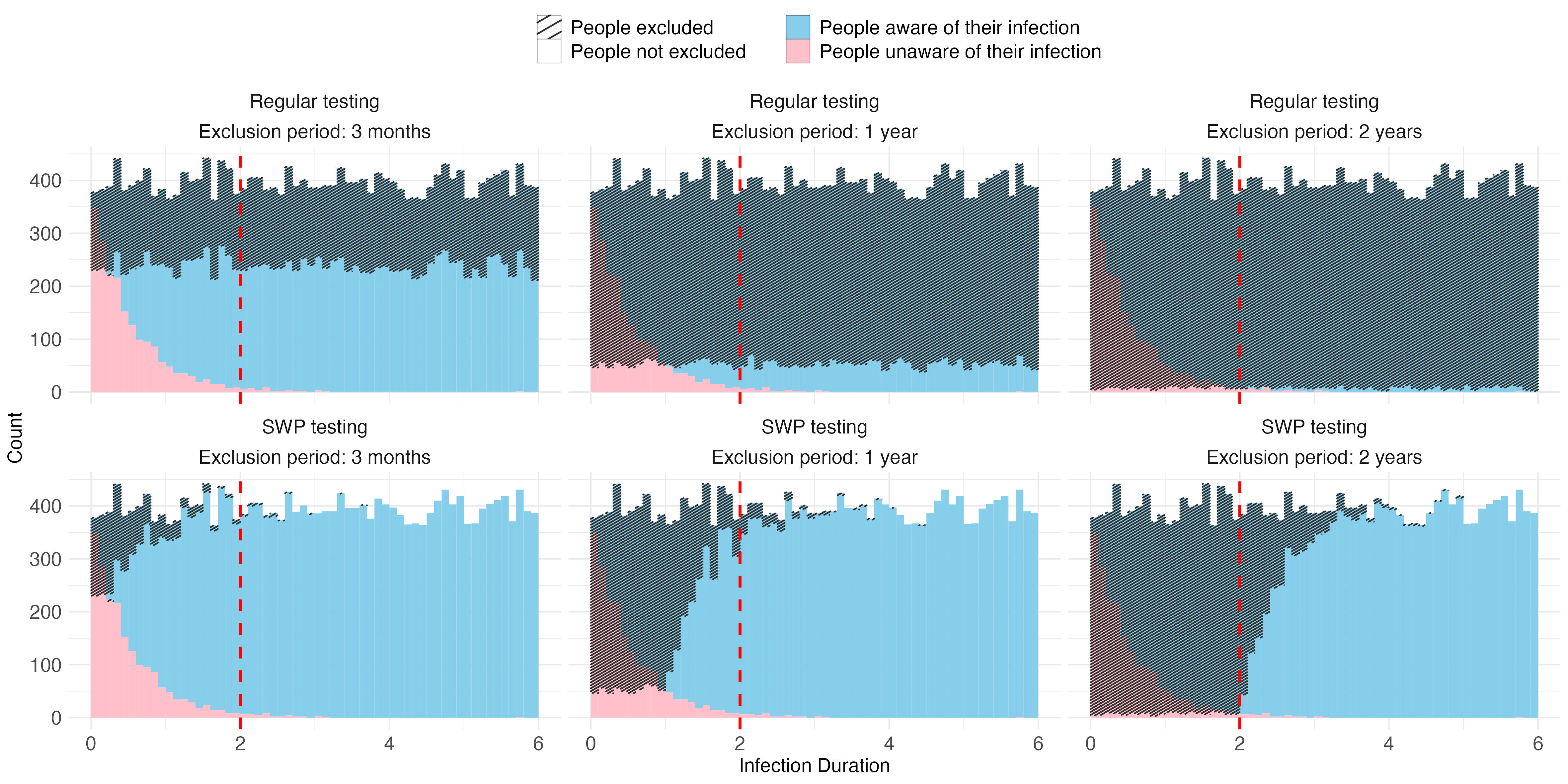}
    \caption{Composition of the HIV-infected population, broken down by their knowledge of their HIV status (color), their infection duration (x-axis), and whether they are excluded from the recency testing sample based on an testing-based criteria (pattern). The pink area represents the distribution of individuals aware of their HIV infections ($f_0(u,c)$), while the blue area represents the distribution of individuals unaware of their HIV infections ($f_1(u,c)$) when regular testing or SWP testing is observed with mean testing frequency of 2 tests/year, using a sample size of 50,000 infected individuals. The influence of the testing-based criterion on $f_0(u,c)$ and $f_1(u,c)$ is depicted under testing-based criterion with $c = 0.25$, $c = 1$, and $c=2$ respectively. The red dashed line is at the time of $T^*$.}
    \label{fig:6_plots_theta2}
\end{figure}

\subsection{Diagnostic histogram of mean testing frequency of 2 tests/year (Figure \ref{fig:6_plots_theta2})}
\newpage
\section{Simulation results for sensitivity analysis} \label{Appendix:sensitivity analysis}\label{appendix:sensitivity}

\begin{figure}[h!]
\centering
    \begin{subfigure}[b]{0.9\textwidth}
        \centering
        \includegraphics[width=0.9\textwidth]{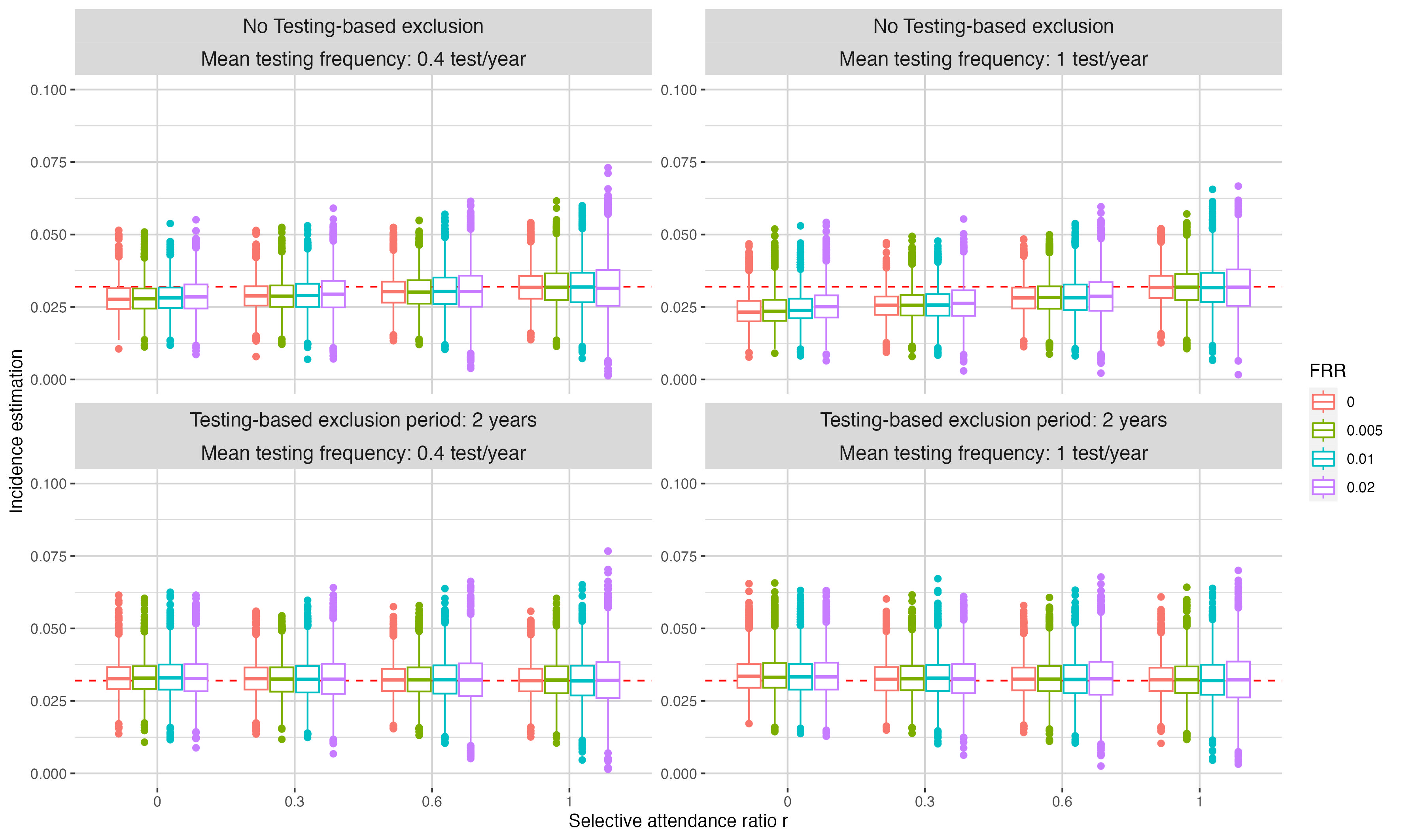}
        \caption{}
    \end{subfigure}
    \hfill
    \begin{subfigure}[b]{0.9\textwidth}
    \centering
    \includegraphics[width=0.9\textwidth]{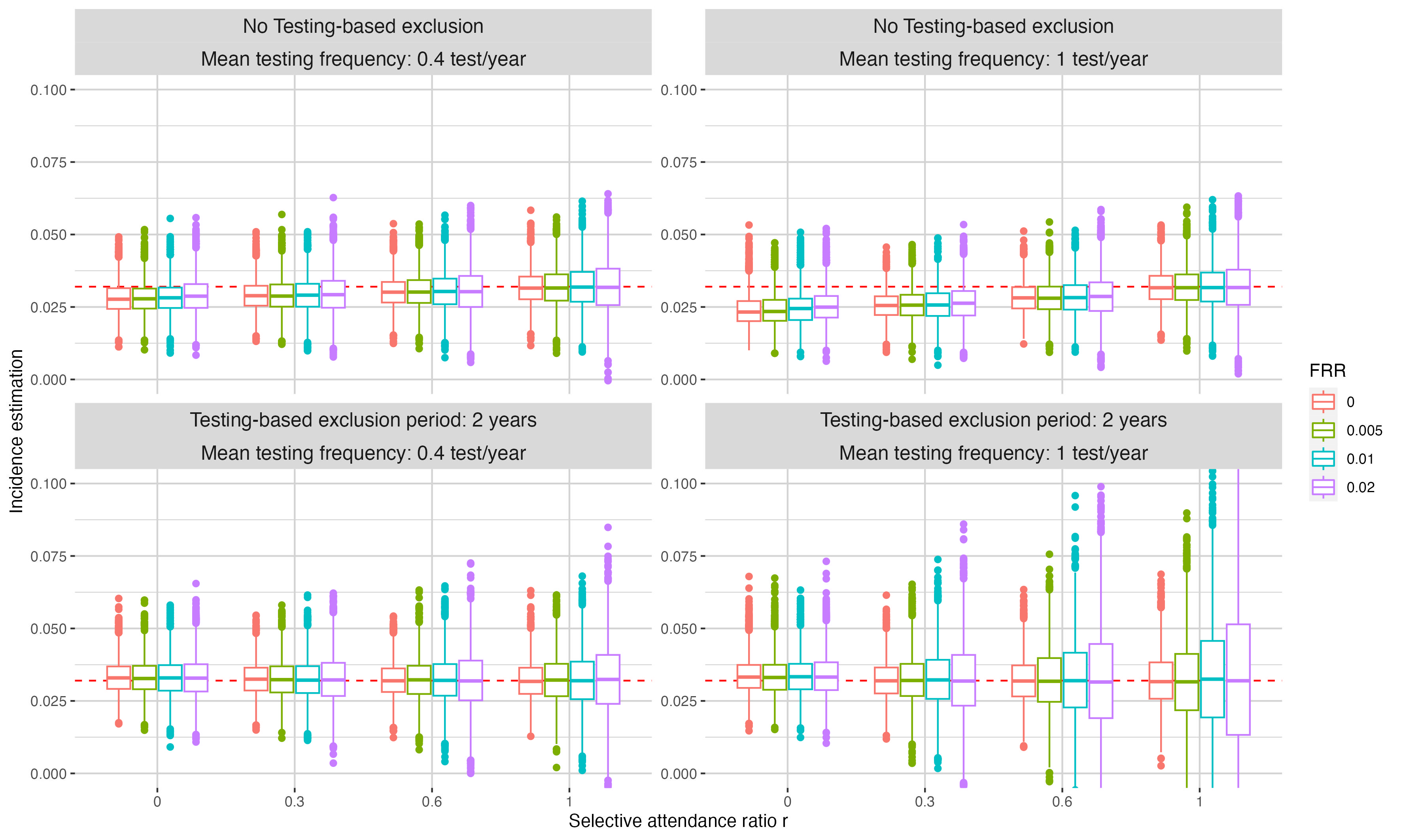}
    \caption{}
    \end{subfigure}
    \caption{
    Performance of incidence estimation under (a) regular testing (Assumption \ref{assumption:id}) and (b) SWP testing (Assumption \ref{assumption:SWP}) with FRR $\beta_{T^*} \in \{0,0.5\%,1\%,2\%\}$. Each subplot shows results for different mean testing frequencies $\theta$, with the x-axis representing the selective attendance ratio $r$ and different colors indicating various testing-based exclusion periods $c$. The true incidence rate is marked by the red line at 0.032.}
    \label{fig:box_frr}
\end{figure}

\subsection{Simulation results of FRR $\beta_{T^*} \neq 0$ (Figure \ref{fig:box_frr})}
\begin{figure}[h]
\centering
    \begin{subfigure}[b]{0.9\textwidth}
        \centering
        \includegraphics[width=\textwidth]{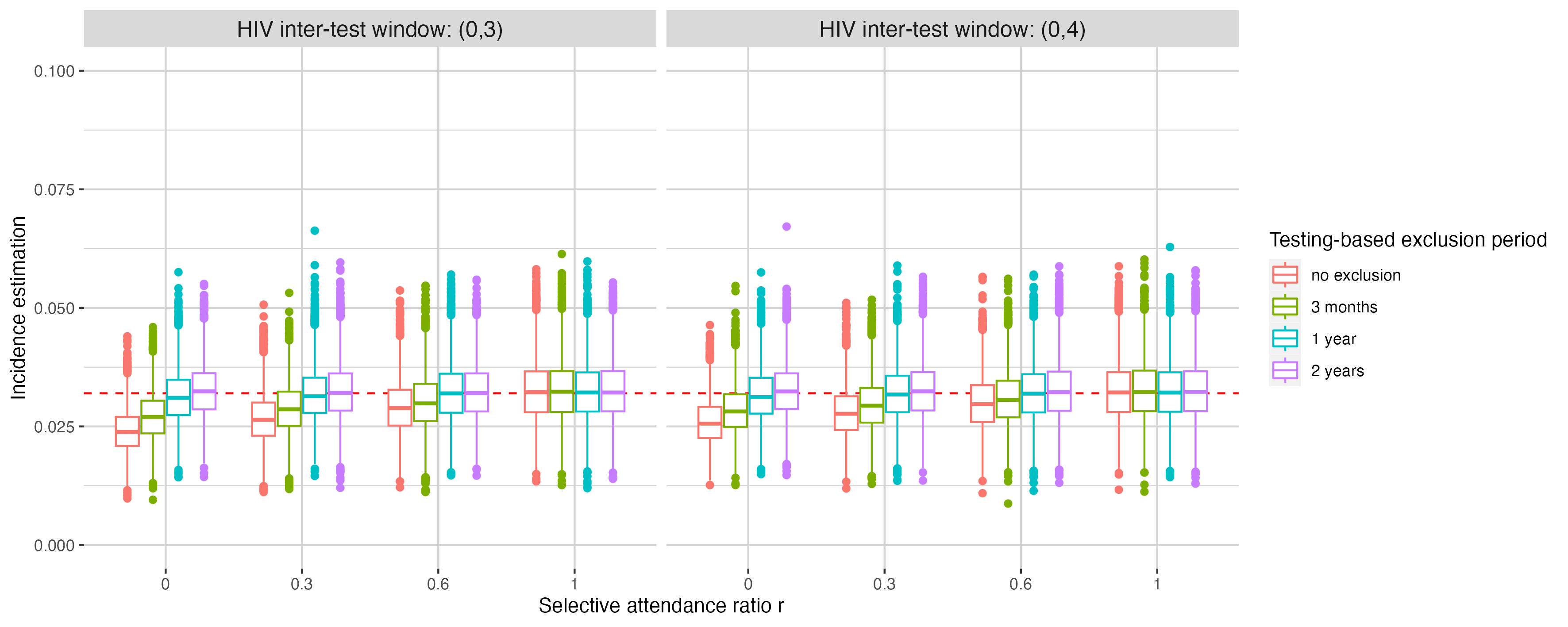}
        \caption{}
    \end{subfigure}
    \hfill
    \begin{subfigure}[b]{0.9\textwidth}
    \centering
    \includegraphics[width=\textwidth]{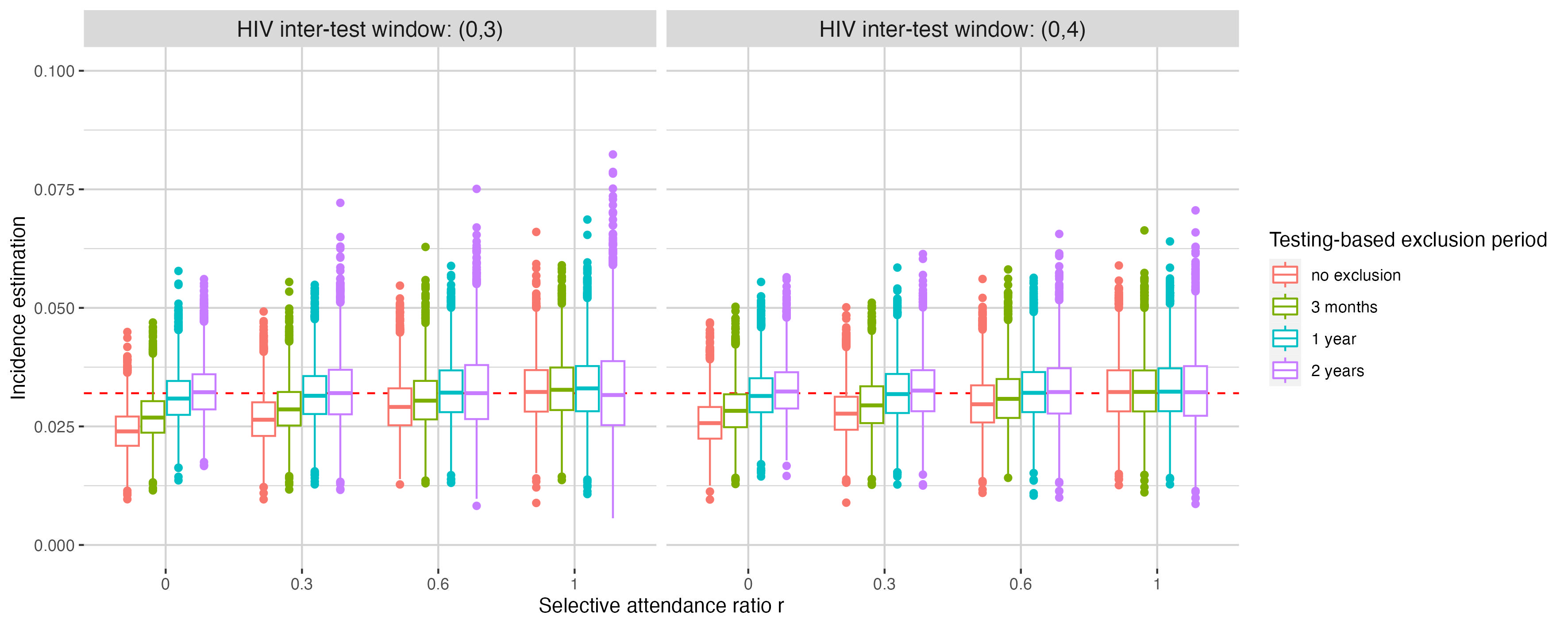}
    \caption{}
    \end{subfigure}
    \caption{Performances of incidence estimation with HIV inter-test time following uniform distribution  under (a) regular testing (Assumption \ref{assumption:id}) and (b) SWP testing (Assumption \ref{assumption:SWP}). Each subplot shows results for different mean testing frequencies $\theta$, with the x-axis representing the selective attendance ratio $r$ and different colors indicating various testing-based exclusion periods $c$. The true incidence rate is marked by the red line at 0.032.
    }
    \label{fig:box_uni}
\end{figure}

\subsection{Simulation results of ,HIV inter-test time following uniform distribution (Figure \ref{fig:box_uni})}
\begin{figure}[h!]
\centering
    \begin{subfigure}[b]{0.9\textwidth}
        \centering
        \includegraphics[width=0.9\textwidth]{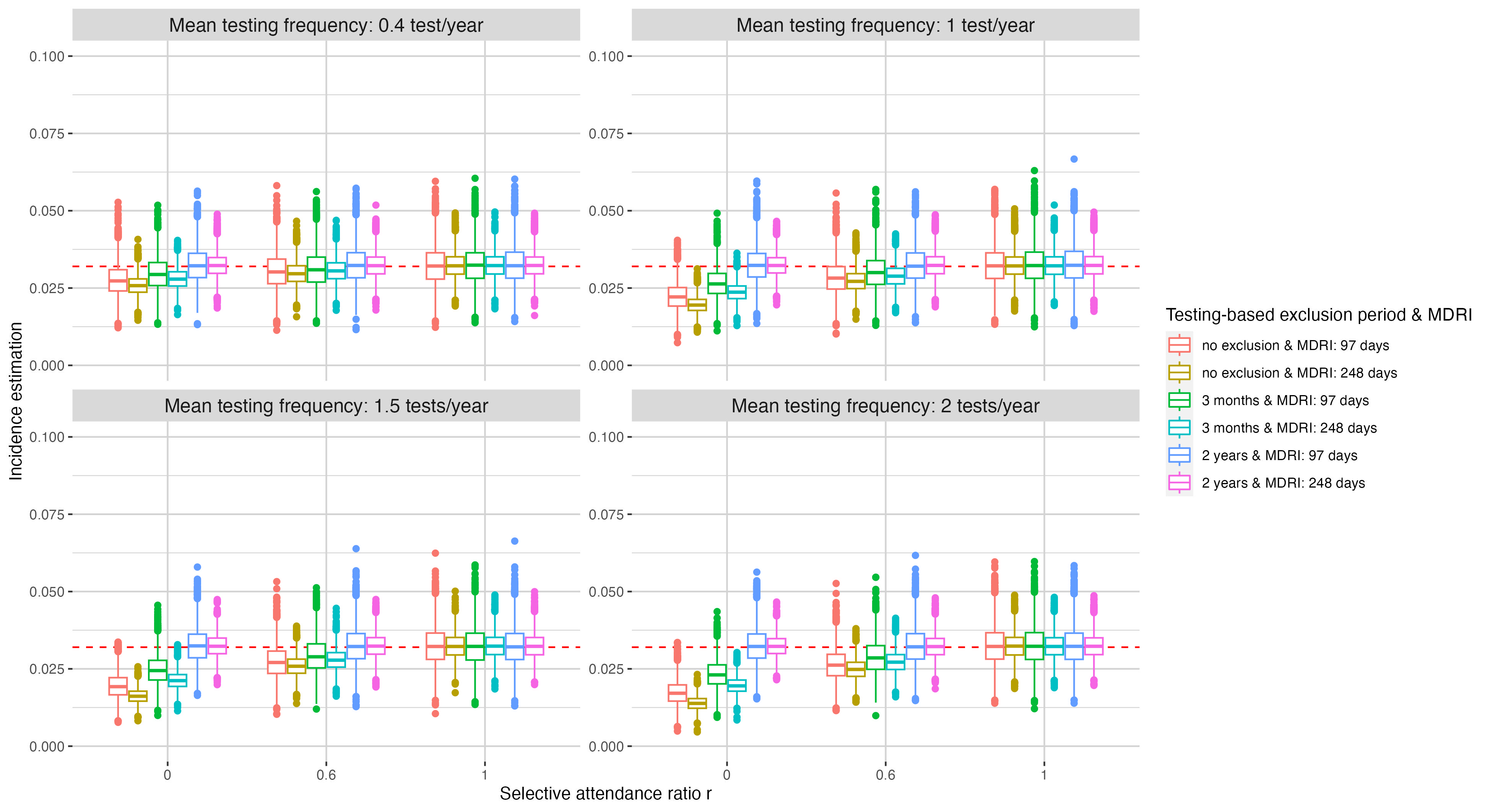}
        \caption{}
    \end{subfigure}
    \hfill
    \begin{subfigure}[b]{1\textwidth}
    \centering
    \includegraphics[width=0.9\textwidth]{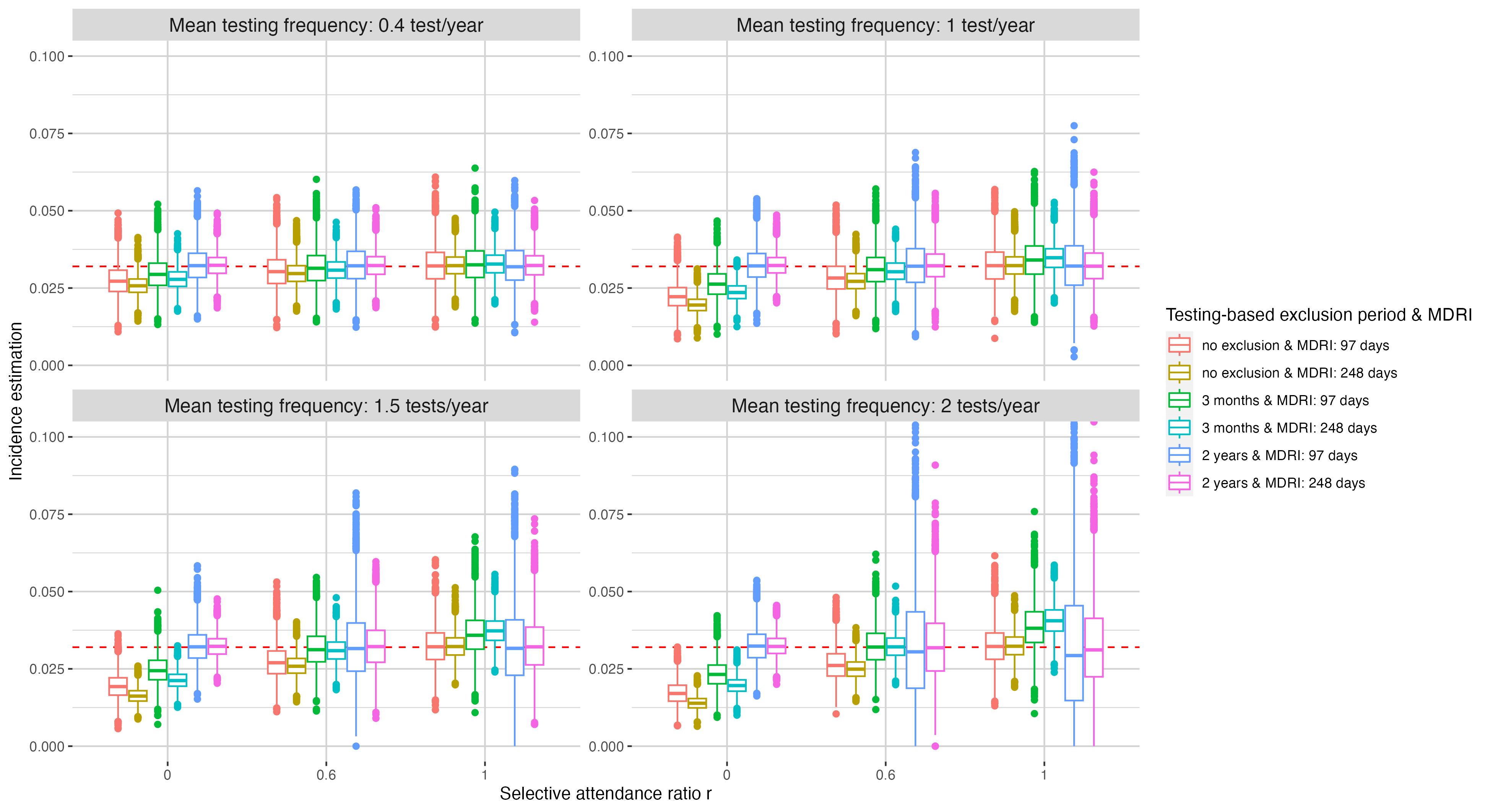}
    \caption{}
    \end{subfigure}
    \caption{Performances of incidence estimation with a longer MDRI (248 days) estimated from external data under (a) regular testing (Assumption \ref{assumption:id}) and (b) SWP testing (Assumption \ref{assumption:SWP}). Each subplot shows results for different mean testing frequencies $\theta$, with the x-axis representing the selective attendance ratio $r$ and different colors indicating various testing-based exclusion periods $c$. The true incidence rate is marked by the red line at 0.032.}
    \label{fig:box_1B}
\end{figure}

\newpage
\subsection{Simulation results of longer MDRI (Figure \ref{fig:box_1B})}

\section{Incidence estimation using effective MDRI when HIV testing history following homogeneous Poisson process} \label{Appendix: effMDRI simulation}
Here we derive the effective MDRI when HIV testing history following homogeneous Poisson process with intensity $\theta$ and negligible FRR ($\beta_{T^*} = 0$). Using the effective MDRI, According to Lemma \ref{lemma:renewal}, $T^{\text{ID}} \sim \text{Exponential}(\theta)$. Therefore, under Assumption \ref{assumption:id},
\begin{align*}
    \Omega_{T,\text{eff}} &= \Omega_{T^*} - (1-r) \frac{\int_0^{T^*} \phi(u) \Pr(T^{\text{ID}} \le u, T^{\text{ID}} > c | U=u,D =1)du}{\Pr(T^{\text{ID}} > c | D = 0)} \\
    &= \Omega_{T^*} - (1-r) \frac{p}{\lambda (1-p)}  \frac{\int_0^{T^*} \phi(u) \Pr(c<T^{\text{ID}} \le u) du}{\Pr(T^{\text{ID}} > c | D = 0)} \\
    &= \Omega_{T^*} - (1-r)  \int_c^{T^*} \phi(u) (1-e^{\theta(c-u)})du
\end{align*}
Given the distribution of $T^{\text{SWP}} $ in (\ref{eq:swp_dist1}) and (\ref{eq:swp_dist2}), the effective MDRI under SWP testing could be written as
\begin{align*}
    \Omega_{T,\text{eff}} &= \frac{\int_0^{T^*} \phi(u)  \{r\Pr(T^{\text{SWP}} > c, T^{\text{SWP}}\le u |U = u, D =1)  + \Pr( T^{\text{SWP}} > c, T^{\text{SWP}}> u | U = u,D =1) \} du}{\Pr(T^{\text{SWP}} > c | D = 0)}\\
    &= \frac{p}{\lambda (1-p)} \frac{r \int_0^{T^*} \phi(u)  \int_c^u f_{T^{SWP}|U}(t) f(u) dtdu   + \int_0^{T^*} \phi(u) \int_{max\{c,u\}}^\infty f_{T^{SWP}|U}(t)f(u) \} dtdu}{\Pr(T^{\text{SWP}} > c | D = 0)} \\
    &= \frac{ r \int_0^{T^*} \phi(u)  \int_c^u f_{T^{SWP}|U}(t) dtdu   + \int_0^{c} \phi(u) \int_{c}^\infty f_{T^{SWP}|U}(t) + \int_c^{T^*} \phi(u) \int_{u}^\infty f_{T^{SWP}|U}(t) \} dtdu}{\Pr(T^{\text{SWP}} > c | D_i = 0)} \\
     &= e^{\theta c} \{ r \int_0^{T^*} \phi(u)  \int_c^u\theta e^{-\theta(u-t)} dt du   + \int_0^{c} \phi(u) \int_{c}^\infty  \theta e^{-\theta t} dtdu+ \int_c^{T^*} \phi(u) \int_{u}^\infty \theta e^{-\theta t} dtdu \} \\
     &= e^{\theta c} \{r \int_c^{T^*} \phi(u)  (1-e^{\theta(c-u)}) du   + \int_0^{c} \phi(u) e^{-\theta c}du+ \int_c^{T^*} \phi(u) e^{-\theta u}du \} \\
     &= \Omega_{T^*} - (1-r e^{\theta c}) \int_c^{T^*} \phi(u) (1-e^{\theta (c-u)}) du.
\end{align*}

\end{document}